# Ehrenfest Dynamics with Spontaneous Localization

Anderson A. Tomaz,[1] Rafael S. Mattos,[1] Saikat Mukherjee,[2] Mario Barbatti[1,3]*

*¹ Aix Marseille University, CNRS, ICR, 13397 Marseille, France*

*² Faculty of Chemistry, Nicolaus Copernicus University in Torun, 87100 Torun, Poland*

*³ Institut Universitaire de France, 75231 Paris, France*

*\* Corresponding author: mario.barbatti@univ-amu.fr; www.barbatti.org*

## Abstract

We propose Ehrenfest Dynamics with Spontaneous Localization (SLED), a decoherence-corrected extension of Ehrenfest dynamics based on the Gisin-Percival quantum-state diffusion (QSD) equation. In SLED, the electronic wavefunction evolves stochastically in the adiabatic energy basis, producing trajectory-level localization. The trajectory ensemble reproduces a Lindblad-type propagation of the reduced electronic density matrix. This approach ensures linearity, trace preservation, and complete positivity, providing a physically consistent alternative to *ad hoc* decoherence corrections commonly adopted in mixed quantum-classical methods. Benchmark simulations on one-dimensional Tully models and multidimensional spin-boson Hamiltonians demonstrate that SLED reproduces electronic populations and captures the essential features of coherence decay. The tests, however, also revealed that accurate treatment will require generalizing the localization kernel controlling the electron-nucleus coupling strength, from a constant into a function of time and phase space coordinates. SLED is implemented in the newly developed Skitten program and will be integrated into Newton-X. While the present work serves as a proof of concept, SLED establishes a rigorous and extensible framework that bridges mixed quantum-classical dynamics with open quantum system theory.

## 1. Introduction

The development of mixed quantum-classical dynamics (MQCD) methods for simulating nonadiabatic phenomena has significantly advanced research in the field.[1] These methods dramatically reduce the complexity and computational cost of solving the time-dependent Schrödinger equation (TDSE) for electronically excited molecular systems by treating light particles (usually electrons) quantum mechanically while describing heavy particles (nuclei) via classical mechanics. Nonadiabatic effects, which describe electronic population transfer between states, are included via algorithms that couple the quantum and classical subsystems. Among such coupling algorithms, the most frequently adopted are surface hopping (which stochastically distributes the classical trajectories among electronic states) and mean field (which propagates classical trajectories on a quantum-weighted averaged potential energy surface).[2]

Much of the popularity of MQCD methods is due to the independent trajectory approximation often adopted. This approximation neglects global quantum correlations, thereby restricting the quantum information to that which is locally calculated at the classical nuclear positions. Thus, independent trajectory methods are ideally suited for execution through on-the-fly techniques, where electronic properties are calculated simultaneously with the trajectories, thereby alleviating the burden of computing global potential energy surfaces before dynamics simulations.

However, the independent trajectory approximation may lead to a faulty description of electronic quantum coherences because information on dephasing between different nuclear wave packet components is unavailable, as recognized long ago.[3] Solving this problem has led to an active research program aimed at developing coherence-corrected MQCD methodologies.[4-14] Nevertheless, many of these methods propose *ad hoc* corrections without guaranteeing they will generally work. For instance, the most used decoherence correction in surface hopping, the simplified decay of mixing (SDM),[12] despite its general success, fails for one of the most basic test cases, the 1-D Tully 2 model (see Figure 3, which will be discussed later).

We aim to develop an MQCD method that still relies on classical nuclear trajectories but that is anchored on a more accurate quantum treatment. We began with the Gisin-Percival quantum-state diffusion (QSD) equation[15] to build a mean-field approach, which naturally incorporates decoherence under the independent trajectory approximation. This treatment led to the Ehrenfest Dynamics with Spontaneous Localization (SLED), which we introduce here. It is based on the standard Ehrenfest dynamics but with the electronic equations of motion following the QSD equation instead of the TDSE.

The QSD equation was proposed in the context of open quantum systems, aiming to describe the stochastic evolution of a vector state that corresponds to a deterministic Lindblad master equation for the system's reduced density (see Supplementary Material, SM-1).[15, 16] It adds a stochastic term to the TDSE so that each time we solve it starting from the same initial conditions (one realization), the QSD equation converges to a different result. These results are the eigenvalues of the operators composing the stochastic term and are distributed with the Born rule over an ensemble of realizations.

Beyond its initial application in quantum diffusion, the spontaneous localization in the QSD equation provided a framework to describe quantum state collapse. It has been widely used in objective collapse theories that modify the TDSE to make collapse a dynamic process.[17] However, this feature can be adapted to simulate decoherence in MQCD. (For the distinction between collapse and decoherence, see Ref. [18]) In an ensemble of independent trajectories composing a classical wave packet,[19] decoherence shows as localization on one eigenvalue, recovering the reduced electronic density matrix corresponding to an improper mixed state over an ensemble of trajectories. Thus, we can use the QSD equation to simulate many trajectories with the desired decoherence treatment.

Another feature of the QSD equation that makes it an excellent starting point for MQCD approaches is that it is constrained to yield valid quantum states. In addition to the stochastic contribution, the QSD equation introduces a new deterministic term to the TDSE. This term ensures that the ensemble average over realizations reduces to the Lindblad equation (see SM-1). This recovery is crucial because it guarantees that the ensemble density matrix evolves in a *linear*, *trace-preserving*, and *completely positive* manner (CPTP map), properties generally required for physical consistency,[20] and usually neglected in MQCD. Linearity prevents superluminal signaling,[21] trace preservation ensures normalization, and complete positivity secures valid predictions even when the system is entangled with external environments. Complete positivity, however, is not an absolute physical necessity; it can be relaxed in models involving initial system-environment correlations or strong coupling.[22, 23] Both the new deterministic and the stochastic terms in the QSD equation stem from a non-Hermitian effective Hamiltonian, resulting in a dissipative time evolution characteristic of open quantum systems.

A final QSD equation feature that makes it potentially interesting for nonadiabatic dynamics is that it is intrinsically flexible concerning the choice of pointer basis, whose states are least affected by the environment, remaining well-defined and stable over time.[24] Here, we will exclusively adopt adiabatic electronic states as the pointer basis. However, it is beneficial to start developing a method that inherently contains such flexibility to be explored in the future in simulations evolving on other pointer bases, such as site-localized electronic states in exciton transport.[25]

We developed SLED by starting from Ehrenfest instead of surface hopping because the former consistently treats the classical and quantum subsystems, with electrons and nuclei evolving on the same state superposition, more effectively. Surface hopping, in turn, is fundamentally inconsistent with the nuclei always moving on a single potential energy surface, while the electrons may be (and generally are) in a quantum superposition of states. For surface hopping, this inconsistency remains until the end of the trajectory (or until decoherence is achieved when decoherence corrections are added). Recognizing this fundamental advantage of the Ehrenfest approach has motivated the development of several methodologies.[26-28]

SLED shares a few elements with previously proposed methods. Prezhdo proposed a mean field approach based on a Hermitian version of the QSD equation.[29] However, it does not recover the Lindblad equation in the ensemble density. Our method sticks to the original non-Hermitian version of the QSD equation, inheriting all its properties. Moreover, it also diverges from Prezhdo's approach because its non-Hermitian character required entirely distinct force calculations and the introduction of energy conservation restrictions.

Truhlar and co-authors have championed the use of decoherence-corrected mean-field dynamics with stochastic jumps, utilizing energy-based decoherence times.[11, 27] Mean field with stochastic jumps is also the strategy Echer and Levine adopted.[28] Nevertheless, their method targets dense state manifolds where coherence may persist within subspaces. Moreover, they take decoherence times proportional to force differences. Gao and Thiel proposed a surface-hopping approach in which decoherence was considered via a non-Hermitian term.[30] All those methods are unrelated to the QSD equation.

This paper is the first step of an extensive research program. We present the basis of the SLED method and results for diverse test cases. However, this presentation should be considered a proof of concept. We are still working to make the approach more flexible (so it can be used with different electronic structure methods and pointer bases) and self-consistent (eliminating an arbitrary constant inherited from Gisin-Percival's approach). Despite this prototypical character, we believe that SLED and its results for model Hamiltonians are already interesting enough for the MQCD community, which motivates this paper.

## 2. Decoherence

### 2.1. Defining Decoherence

Coherence evolution is the central motivation for SLED development. For this reason, this section provides a detailed discussion of decoherence in the MQCD context. We recommend the excellent reviews by Shu and Truhlar[31] and Schultz *et al.*[25] for an extended discussion of decoherence in the molecular context. For more general aspects, see also the authoritative book by Schlosshauer.[32]

Decoherence is a feature of open quantum systems, where quantum coherences are dispersed into the environment through interaction. To understand how it occurs, consider a general bipartite quantum state split between system ($S$) and environment ($E$):

$$\left|\Psi\left(t\right)\right\rangle = \sum_{N} A_{N}\left(t\right)\left|S_{N}\right\rangle\left|E_{N}\right\rangle, \tag{1}$$

with $\sum_{N}\left|A_{N}\right|^{2} = 1$ and we will drop the dependence on $t$ to simplify the notation. The density matrix is

$$\begin{aligned} \rho_{SE}\left(t\right) &\equiv \left|\Psi\right\rangle\left\langle\Psi\right| \\ &= \sum_{M,N} A_{M}^{*} A_{N}\left|S_{N}\right\rangle\left\langle S_{M}\right|\left|E_{N}\right\rangle\left\langle E_{M}\right|. \end{aligned} \tag{2}$$

The reduced density matrix of the system is

$$\begin{aligned} \rho_{S}\left(t\right) &= Tr_{E}\left[\rho_{SE}\right] \\ &= \sum_{M,N} A_{M}^{*} A_{N}\left\langle E_{M}\middle|E_{N}\right\rangle\left|S_{N}\right\rangle\left\langle S_{M}\right|. \end{aligned} \tag{3}$$

The diagonal elements $\rho_{S,LL} = \left|A_{L}\right|^{2}$ are the system's populations of state $L$, and the off-diagonal elements $\rho_{S,KL} = A_{K}^{*} A_{L}\left\langle E_{K}\middle|E_{L}\right\rangle = \rho_{S,LK}^{*}$ are the system's coherences between states $K$ and $L$.

Initially, we have no reason to suppose that the environment states form an orthogonal basis, meaning that coherence can be non-null. However, given enough time, the environment states tend to become orthogonal, and the system's coherence approaches zero. This process is called decoherence,[32] occurring within a characteristic decoherence time, $\tau_{D}$. After decoherence, the reduced density matrix of the system is diagonal, $\rho_{S,KL}\left(t \gg \tau_{D}\right) = \delta_{KL}$.

Although decoherence can occur in any basis, decoherence theory identifies a preferred one, the pointer basis. The emergence of these states, known as environment-induced superselection (einselection), is governed by the structure of the total Hamiltonian. Two limiting cases are particularly relevant.[33] When the system-environment interaction dominates, the pointer states tend to be eigenstates of the interaction Hamiltonian. Conversely, when the system's self-Hamiltonian dominates and the environment evolves adiabatically—i.e., on timescales much slower than those of the system—the pointer states align with the system's energy eigenstates. This regime corresponds to what Paz and Zurek termed the quantum limit of decoherence.[33]

## 2.2. Electronic Decoherence in Nonadiabatic Dynamics

Quantum coherences between electronic excited states can be dispersed to the nuclei, which act as an environment interacting with the electrons. In a molecule, the electronic frequencies are much bigger than the nuclear frequencies. Under this adiabatic nuclear environment,[33] the electronic energy eigenstates emerge as pointer states of the electronic system entangled with the nuclear vibrations. This realization motivates writing the quantum molecular state in terms of electronic energy eigenstates $|N\rangle$ as

$$\left|\Theta(t)\right\rangle = \sum_N |N\rangle \left|\Omega_N(t)\right\rangle, \tag{4}$$

where $\left|\Omega_N(t)\right\rangle$ are nuclear states. Eq. (4) is the Born-Huang approximation,[34] the starting points for many of the nonadiabatic dynamics approaches available.[35]

Following the previous discussion (see eq. (3)), the electronic population of state $L$ is

$$\rho_{e,LL} = \left\langle \Omega_L(t) \middle| \Omega_L(t) \right\rangle = \int d\mathbf{R} \left| \Omega_L(\mathbf{R},t) \right|^2, \tag{5}$$

where $\Omega_L(\mathbf{R},t)$ is the nuclear wave packet in state $L$. The coherence between states $L$ and $K$ is

$$\rho_{e,KL} = \left\langle \Omega_K(t) \middle| \Omega_L(t) \right\rangle = \int d\mathbf{R}\, \Omega_K(\mathbf{R},t)^* \, \Omega_L(\mathbf{R},t) = \rho_{e,LK}^*. \tag{6}$$

This last expression unveils that the source of electronic decoherence is modulated by the dephasing between components of the nuclear wave packet in different electronic states.

Indeed, when the potential energy surfaces are parallel and near degenerate, the electronic coherence tends to persist for long times.[36]

## 2.3. Electronic Decoherence in Mixed Quantum-Classical Dynamics

The MQCD decoherence problem has two aspects. The first one, the *consistency problem*, is the divergence between occupation (the fraction of trajectories in each state) and the electronic mean population.[12] The consistency problem, which particularly impacts the fewest-switches surface hopping (FSSH),[37] is simple to tackle. Any decoherence correction, even a simple instantaneous coherence-erasing after hopping, solves it. Indeed, in cases where the molecule does not return to the crossing region after hopping, we do not even need to worry about it if we base the statistical analysis exclusively on the occupation.

The second decoherence issue, the *persistency problem*, is more challenging to address and may affect mean-field approaches and other versions of surface hopping beyond FSSH. It occurs when the molecule reaches the nonadiabatic interaction region between the same states multiple times during the trajectory. In such a case, the coherence built during one passage may impact the nonadiabatic transition probability during the second passage. Therefore, we must account for how long the coherence persists to make reliable predictions.

To understand the origin of these problems in specific MQCD methods, such as Ehrenfest and FSSH,[2] we can explicitly derive the decoherence following the approach from the previous section. We assign to each independent trajectory $k$ a time-dependent electronic state of the form

$$\left|\Psi^{(k)}(t)\right\rangle = \sum_N A_N^{(k)}(t)\left|N(t)\right\rangle, \tag{7}$$

where $\left|N(t)\right\rangle$ is the electronic energy eigenstate $N$ determined at the nuclear geometry $\mathbf{R}^{(k)}(t)$. Thus, the ensemble of trajectories yields an electronic density

$$\rho_e^{(MQCD)}(t) \approx \sum_{N,M}\left(\frac{1}{N_{traj}}\sum_j A_N^{(k)}(t)A_M^{(k)}(t)^*\right)\left|N\right\rangle\left\langle M\right|, \tag{8}$$

where we can drop the basis dependence on time because the ensemble spans (in principle) the entire nuclear configuration space. This expression defines the MQCD electronic populations

$$\rho_{e,LL}^{(MQCD)} = N_{traj}^{-1} \sum_j \left| A_L^{(j)}(t) \right|^2 \tag{9}$$

and coherences

$$\rho_{e,KL}^{(MQCD)} = N_{traj}^{-1} \sum_j A_L^{(j)}(t) A_K^{(j)}(t)^* = \rho_{e,KL}^{(MQCD)*}. \tag{10}$$

Comparing the quantum coherence [Eq. (6)] and the MQCD coherence in Eq. (10) highlights a crucial problem with the latter: it does not contain any information about the nuclear overlap dephasing, and therefore, it does not lose coherence.[38] Moreover, the only straightforward way to impose decoherence in such an ensemble density is by damping the coefficient $A_L(t)$ for one of the states in each trajectory. This is precisely how it is done in the many proposed methods to introduce decoherence corrections in MQCD.[5, 14] Some of these methods include damping terms in the equation of the motion,[11, 27, 29, 30, 39] while in others, the coefficients are damped separately from the integration.[7, 12, 28, 40-43] Central to all those methods is defining the decoherence rate. In this aspect, they are divided into two types: methods estimating nuclear overlaps, which yield rates proportional to force differences[5, 7, 28, 41] or energy gap fluctuations,[13] and methods that estimate the decoherence rate from the time-energy uncertainty principle,[11, 12, 27] yielding rates inversely proportional to energy gaps.

## 3. Ehrenfest Dynamics

### 3.1. Time-Dependent Self-Consistent Field

In Ehrenfest dynamics, the electrons evolve according to

$$\frac{d}{dt} \left| \Psi(t) \right\rangle = -\frac{i}{\hbar} \hat{H} \left| \Psi(t) \right\rangle, \tag{11}$$

and the expected value of the nuclear wave packet follows

$$\frac{d^2\mathbf{R}_\alpha}{dt^2} = -\frac{1}{M_\alpha}\left\langle \Psi(t)\left|\nabla\hat{H}\right|\Psi(t)\right\rangle_\mathbf{r} \tag{12}$$

for each nucleus $\alpha$. In these equations, $\hat{H}$ is the electronic Hamiltonian and $\left|\Psi(t)\right\rangle$ is the time-dependent electronic quantum state under the condition that the nuclei are at position $\mathbf{R}$. $\mathbf{r}$ represents the ensemble of electronic coordinates, and the subindex $\mathbf{r}$ indicates integration over this variable.

These equations of motion (EOM) arise in two steps. First, we factorize the molecular wave function $\Theta(\mathbf{r}, \mathbf{R}, t)$ as

$$\Theta(\mathbf{r},\mathbf{R},t) = \Omega(\mathbf{R},t)\Psi(\mathbf{r},t)e^{i\phi(t)}, \tag{13}$$

where $\Omega(\mathbf{R}, t)$ is the nuclear component.[44] The phase factor is

$$\phi(t) = \frac{1}{\hbar}\int_0^t \left\langle \Omega(t')\Psi(t')\left|\hat{H}\right|\Omega(t')\Psi(t')\right\rangle_{\mathbf{rR}} dt'. \tag{14}$$

Inserting Ansatz (13) into the TDSE decouples it into two effective Schrödinger equations,[44] one for the electrons (Eq. (11)), and another for the nuclei:

$$\frac{d}{dt}\left|\Omega(t)\right\rangle = -\frac{i}{\hbar}\left(\hat{T}_N + \left\langle \Psi(t)\left|\hat{H}\right|\Psi(t)\right\rangle_\mathbf{r}\right)\left|\Omega(t)\right\rangle. \tag{15}$$

Eqs. (11) and (15) are known as the time-dependent self-consistent field (TDSCF) formulation.

The second step to get the Ehrenfest EOMs consists of using the Ehrenfest theorem[45]

$$\frac{d}{dt}\left\langle \hat{A}\right\rangle = \frac{1}{i\hbar}\left\langle \left[\hat{A},\hat{\mathcal{H}}\right]\right\rangle + \left\langle \frac{\partial\hat{A}}{\partial t}\right\rangle \tag{16}$$

to predict the time derivatives of the expected values of the nuclear coordinate position and momentum operators ($\hat{\mathbf{R}}$ and $\hat{\mathbf{P}}$) for the Hamiltonian $\hat{\mathcal{H}} = \hat{T}_N + \left\langle \Psi(t)\left|\hat{H}\right|\Psi(t)\right\rangle_\mathbf{r}$. This procedure directly leads to Newton's nuclear equation [Eq. (12)].

This derivation differs from the well-known Tully's derivation of the mean-field EOMs[2, 44] (based on Micha's work[46]), which takes the semiclassical limit of Eq. (15). Tully-Micha's procedure leads to the classical force $\mathbf{F}_\alpha = -\nabla \left\langle \Psi(t) \middle| \hat{H} \middle| \Psi(t) \right\rangle_\mathbf{r}$ instead of the Ehrenfest force $\mathbf{F}_\alpha = -\left\langle \Psi(t) \middle| \nabla \hat{H} \middle| \Psi(t) \right\rangle_\mathbf{r}$. Tully argues that both treatments are equivalent.[44] Nevertheless, this equivalence is rigorously true only for one-dimensional systems. Moreover, we see no practical way to compute the Tully-Micha force when expanding the wavefunction on an adiabatic representation. Therefore, we will stick to the Ehrenfest formulation in Eq. (12), which can be computed as explained in Section 3.3.

There is also a more profound conceptual difference between the two derivations. As mentioned, Eq. (12) describes the motion of the expected value of the nuclear wave packet. On the other hand, the Tully-Micha treatment describes the entire wave packet as an ensemble of independent classical trajectories. This difference is reconciled if the initial conditions for trajectories in the Ehrenfest approach are sampled, supposing a mixed state whose density distribution resembles that of the coherent wave packet.

## 3.2. Quantum Electronic Evolution in the Adiabatic Basis

In Ehrenfest dynamics, the electronic time evolution is determined by the Schrödinger equation [Eq. (11)]. Consider the moving basis defined as

$$\hat{H}(t)\left|N(t)\right\rangle = E_N(t)\left|N(t)\right\rangle, \tag{17}$$

Where the dependence of the basis on time arises from its parametric dependence on the nuclear geometry $\left|N(t)\right\rangle = \left|N\big(\mathbf{R}(t)\big)\right\rangle$ defined in Eq. (12). We can write the electronic state $\left|\Psi(t)\right\rangle$ as

$$\left|\Psi(t)\right\rangle = \sum_N A_N(t) e^{-i\gamma_N(t)} \left|N(t)\right\rangle, \tag{18}$$

where the phase

$$\gamma_N(t) = \frac{1}{\hbar} \int_0^t E_N(t') dt' \tag{19}$$

is adopted to simplify the final result.

Replacing Eq. (18) in (11) and projecting it on one of the electronic states renders

$$\frac{dA_M(t)}{dt} = -\sum_{N \neq M} e^{-i\gamma_{NM}(t)} \sigma_{MN}(t) A_N(t), \tag{20}$$

where

$$\sigma_{MN}(t) = \left\langle M(t) \middle| \frac{dN(t)}{dt} \right\rangle \tag{21}$$

is the time-derivative nonadiabatic coupling, and

$$\gamma_{NM}(t) = \gamma_N(t) - \gamma_M(t). \tag{22}$$

## 3.3. Classical Nuclear Evolution in the Adiabatic Basis

In Ehrenfest dynamics, a nucleus $\alpha$ moves according to Newton's equation [Eq. (12)]. The gradient term can be written as

$$\left\langle \Psi(t) \middle| \nabla \hat{H} \middle| \Psi(t) \right\rangle = \nabla \left\langle \Psi(t) \middle| \hat{H} \middle| \Psi(t) \right\rangle - \left\langle \nabla\Psi(t) \middle| \hat{H} \middle| \Psi(t) \right\rangle - \left\langle \Psi(t) \middle| \hat{H} \middle| \nabla\Psi(t) \right\rangle, \tag{23}$$

where we dropped the integration index to simplify the notation. With (18), we obtain

$$\begin{aligned}
\left\langle \Psi(t) \middle| \nabla \hat{H} \middle| \Psi(t) \right\rangle &= \sum_N |A_N(t)|^2 \nabla E_N(t) \\
&+ \sum_{N, M \neq N} A_M(t)^* A_N(t) e^{-i\gamma_{NM}(t)} \left( E_N(t) - E_M(t) \right) \mathbf{d}_{MN}(t),
\end{aligned} \tag{24}$$

where $\mathbf{d}_{MN} \equiv \int d\mathbf{r}\, \varphi_M(\mathbf{r};\mathbf{R})^* \nabla \varphi_N(\mathbf{r};\mathbf{R})$ is the nonadiabatic coupling vector written in terms of the electronic integral of the adiabatic wave function $\varphi_N(\mathbf{r};\mathbf{R}) = \langle \mathbf{r} | N \rangle$ for state $N$ at nuclear geometry $\mathbf{R}$. In this derivation, we used $\mathbf{d}_{NM} = -\mathbf{d}_{MN}$, which follows directly from $\nabla \int d\mathbf{r}\, \varphi_M(\mathbf{r};\mathbf{R})^* \varphi_N(\mathbf{r};\mathbf{R}) = 0$. The previous expression can be more conveniently expressed as

$$\left\langle \Psi(t) \left| \nabla \hat{H} \right| \Psi(t) \right\rangle = \sum_N \left| A_N(t) \right|^2 \nabla E_N(t)$$
$$- 2 \sum_{N, M > N} \mathrm{Re}\left[ A_M(t) A_N(t)^* e^{-i\gamma_{MN}(t)} \right] \left( E_M(t) - E_N(t) \right) \mathbf{d}_{MN}(t). \tag{25}$$

## 4. Ehrenfest Dynamics with Spontaneous Localization

### 4.1. The Gisin-Percival QSD Equation

According to Gisin and Percival, a quantum state can be written as

$$\left| \Psi(t) \right\rangle = \frac{\left| \Phi(t) \right\rangle}{\left\| \Phi(t) \right\|}, \tag{26}$$

where $\left| \Phi(t) \right\rangle$ follows the QSD equation[21, 47]

$$d\left| \Phi \right\rangle = -\frac{i}{\hbar} \hat{H} \left| \Phi \right\rangle dt - \frac{1}{2} \sum_n \left( \hat{L}_n^\dagger - \left\langle \hat{L}_n^\dagger \right\rangle \right) \left( \hat{L}_n - \left\langle \hat{L}_n \right\rangle \right) \left| \Phi \right\rangle dt + \sum_n \left( \hat{L}_n - \left\langle \hat{L}_n \right\rangle \right) \left| \Phi \right\rangle dW_n, \tag{27}$$

where $\left\langle L_n \right\rangle = \left\langle \Phi \left| \hat{L}_n \right| \Phi \right\rangle / \left\langle \Phi | \Phi \right\rangle$. If only the first term on the right-hand side of Eq. (27) is retained, we have the conventional Schrödinger evolution. The two additional terms cause the equation to localize spontaneously in one of the eigenvectors of $\hat{L}_n$. The last term makes this localization random, whereas the second term ensures that the evolution of the density computed over an ensemble of stochastic realizations of Eq. (27) is linear (SM-1).

The state localization is stochastic and controlled by a complex Wiener process $dW_n$ with[15]

$$\begin{aligned}
&\left\langle dW_n \right\rangle = 0, \\
&\left\langle \mathrm{Re}(dW_m) \mathrm{Re}(dW_n) \right\rangle = \left\langle \mathrm{Im}(dW_m) \mathrm{Im}(dW_n) \right\rangle = \delta_{mn} dt, \\
&\left\langle \mathrm{Re}(dW_m) \mathrm{Im}(dW_n) \right\rangle = 0.
\end{aligned} \tag{28}$$

The statistic over multiple realizations of Eq. (27) follows the Born rule. Due to the dissipative character of the $\hat{L}_n$-terms in this equation, the norm of $\left|\Phi(t)\right\rangle$ is not conserved, which leads to the introduction of the physically sound normalized state $\left|\Psi(t)\right\rangle$ defined in Eq. (26).

## 4.2. Stochastic Schrödinger Evolution

In a molecule, the electronic frequencies are much bigger than the nuclear frequencies. Under this adiabatic nuclear environment,[33] the electronic energy eigenstates emerge as pointer states of the electronic system entangled with the nuclear vibrations (see Section 2.1). Thus, $\hat{H}$ becomes the natural choice for the localization operator in the QSD equation:

$$\sum_n \hat{L}_n \to \sqrt{\kappa}\hat{H},\qquad(29)$$

where $\kappa$ is a real and positive localization kernel encoding the electron-nuclear coupling strength. In this work, it will be treated as a constant. In Section 7, we analyze the limitations of such treatment.

Suppose the time-dependent electronic quantum state can be written as Eq. (26), where $\left|\Phi(t)\right\rangle$ follows the QSD Equation (27) with the localization operators given by (29):

$$d\left|\Phi(t)\right\rangle = -\frac{i}{\hbar}\hat{H}\left|\Phi(t)\right\rangle dt - \frac{\kappa}{2}\left(\hat{H} - \langle H\rangle\right)^2\left|\Phi(t)\right\rangle dt + \sqrt{\kappa}\left(\hat{H} - \langle H\rangle\right)\left|\Phi(t)\right\rangle dW.\qquad(30)$$

In this equation, $dW$ is the complex Wiener process defined in Eq. (28), and

$$\langle H\rangle = \frac{\left\langle\Phi\middle|\hat{H}\middle|\Phi\right\rangle}{\langle\Phi|\Phi\rangle}.\qquad(31)$$

We can write $\left|\Phi(t)\right\rangle$ in terms of the moving basis $\left\{\left|N(t)\right\rangle\right\}$ like in Ehrenfest dynamics using Eq. (18). However, because $\left|\Phi(t)\right\rangle$ is not normalized, we slightly change the notation, using coefficients $c_N$ instead of $A_N$:

$$\left|\Phi(t)\right\rangle = \sum_N c_N(t) e^{-i\gamma_N(t)} \left|N(t)\right\rangle. \tag{32}$$

The basis is still defined as in Eq. (17), and the phase as in (19).

With these definitions, the equations of motion for the coefficient **c** are

$$dc_M = -\sum_{N \neq M} e^{-i\gamma_{NM}} \sigma_{MN} c_N dt - \frac{\kappa}{2}\left(E_M - \langle E \rangle\right)^2 c_M dt + \sqrt{\kappa}\left(E_M - \langle E \rangle\right)c_M dW_M, \tag{33}$$

where the expected energy value is

$$\left\langle E \right\rangle_t = \frac{\sum_N E_N(t)\left|c_N(t)\right|^2}{\sum_N \left|c_N(t)\right|^2}. \tag{34}$$

We can also write the normalized state $\left|\Psi(t)\right\rangle$ in terms of $\left\{\left|N(t)\right\rangle\right\}$ as

$$\left|\Psi(t)\right\rangle = \sum_N A_N(t) e^{-i\gamma_N(t)} \left|N(t)\right\rangle, \tag{35}$$

implying

$$\mathbf{A}(t) = \frac{\mathbf{c}(t)}{\sum_N \left|c_N(t)\right|^2}. \tag{36}$$

## 4.3. Total Energy Rescaling

The non-Hermitian character of the QSD equation implies that the total energy is not conserved. In SLED, total energy conservation is ensured by rescaling the velocities so that at the end of the time step, the kinetic energy $K^{(0)}$ changes by a quantity $\delta K$ so that

$$\left\langle E(t+\Delta t)\right\rangle + K^{(0)}(t+\Delta t) + \delta K = \left\langle E(t)\right\rangle + K(t). \tag{37}$$

The velocity rescaling of the nucleus $\alpha$ is

$$\mathbf{v}_\alpha = \mathbf{v}_\alpha^{(0)} + \beta \frac{\mathbf{u}_\alpha}{M_\alpha}, \tag{38}$$

where $\mathbf{v}_\alpha^{(0)}$ is the Newtonian velocity at the end of the time step, $\beta$ is a parameter to be determined, and $\mathbf{u}_\alpha$ is the rescaling direction.

Following Herman[48] and Tully,[49] we assume that the meaningful direction for velocity rescaling is that of the nonadiabatic coupling vector. The coupling between multiple pairs of states may play a role in the mean-field evolution. Thus, we propose to rescale velocity in the weighted mean direction

$$\mathbf{u}_\alpha = \frac{\sum\limits_{N,M>N} \sigma_{MN} \mathbf{d}_{MN,\alpha}}{\sum\limits_{N,M>N} \sigma_{MN}}. \tag{39}$$

This direction considers the importance of the coupling between each pair of states to the evolution of the quantum coefficients. If only two states are involved in the nonadiabatic dynamics, $\mathbf{u}$ trivially becomes the nonadiabatic coupling vector between them.

To determine $\beta$, we write[50]

$$\delta K = \frac{1}{2} \sum_\alpha M_\alpha \left( \mathbf{v}_\alpha \right)^2 - \frac{1}{2} \sum_\alpha M_\alpha \left( \mathbf{v}_\alpha^{(0)} \right)^2. \tag{40}$$

Replacing Eq. (38) in (40) gives

$$\delta K = a\beta^2 + b\beta, \tag{41}$$

where

$$\begin{aligned} a &= \frac{1}{2} \sum_\alpha \frac{\left( \mathbf{u}_\alpha \cdot \mathbf{u}_\alpha \right)}{M_\alpha}, \\ b &= \sum_\alpha \mathbf{v}_\alpha^{(0)} \cdot \mathbf{u}_\alpha. \end{aligned} \tag{42}$$

With Eq. (37), we get

$$a\beta^2 + b\beta + \delta\varepsilon_T = 0, \tag{43}$$

where

$$\delta\varepsilon_T = \left[ \left\langle E\left( t + \Delta t \right) \right\rangle + K^{(0)}\left( t + \Delta t \right) \right] - \left[ \left\langle E\left( t \right) \right\rangle + K\left( t \right) \right]. \tag{44}$$

If $\Delta \equiv b^2 - 4a\delta\varepsilon_T \geq 0$, we choose the value $\beta$ that has the most negligible impact on the original velocity:

$$\beta = \begin{cases} \dfrac{-b+\sqrt{\Delta}}{2a}, & \text{if } \left| -b+\sqrt{\Delta} \right| < \left| -b-\sqrt{\Delta} \right| \\ \dfrac{-b-\sqrt{\Delta}}{2a}, & \text{if } \left| -b+\sqrt{\Delta} \right| \geq \left| -b-\sqrt{\Delta} \right| \end{cases}. \tag{45}$$

If, however, $\Delta < 0$, Eq. (43) has no real solution for $\beta$ and total energy conservation cannot be ensured via velocity rescaling. In such a case, which is analogous to frustrated hoppings in surface hopping, we propose rescaling both velocities and quantum coefficients to reach energy conservation. First, we rescale the velocities to the maximum possible amount, which occurs for $\Delta = 0$ and $\beta = -b/2a$:

$$\mathbf{v}_a\left(t+\Delta t\right) = \mathbf{v}_\alpha^{(0)}\left(t\right) - \frac{b}{2a}\frac{\mathbf{u}_\alpha}{M_\alpha}. \tag{46}$$

Then, the remaining difference between the total energies is eliminated by rescaling the electronic populations.

To rescale the electronic populations, we split the $N_s$ states into two groups, the $N_b$ states below and the $N_u$ states above the mean energy. Then, we displace some population from the upper to the bottom states

$$\begin{aligned} \rho_u\left(t+\Delta t\right) &\rightarrow \rho_u\left(t+\Delta t\right) - \mathrm{B}^2\rho_u\left(t+\Delta t\right), \\ \rho_b\left(t+\Delta t\right) &\rightarrow \rho_b\left(t+\Delta t\right) + \mathrm{B}^2\rho_u\left(t+\Delta t\right), \end{aligned} \tag{47}$$

where

$$\rho_x \equiv \sum_{N \in N_s} \left| a_N \right|^2. \tag{48}$$

With these definitions, the populations of the upper states are rescaled as

$$\rho_u = \rho_u^{(0)}\left(1 - \mathrm{B}^2\right), \tag{49}$$

whereas the bottom states go as

$$\rho_b = \rho_b^{(0)}\left(1 + \frac{\rho_b^{(0)}}{\rho_u^{(0)}}B^2\right). \tag{50}$$

The amount of displaced population is proportional to $B^2$, which is chosen to restore energy conservation:

$$\left\langle E(t+\Delta t)\right\rangle + K(t+\Delta t) = \left\langle E(t)\right\rangle + K(t), \tag{51}$$

where

$$\left\langle E(t+\Delta t)\right\rangle = \left(1 + B^2\frac{\rho_u}{\rho_b}\right)\left\langle E(t+\Delta t)\right\rangle_b + \left(1 - B^2\right)\left\langle E(t+\Delta t)\right\rangle_u, \tag{52}$$

with

$$\left\langle E\right\rangle_x \equiv \sum_{N\in N_x}\left|a_N\right|^2 E_N. \tag{53}$$

Inserting Eq. (52) into (51) and solving it for $B^2$ results in

$$B^2 = -\frac{\rho_b(t+\Delta t)\left(\Delta\left\langle E\right\rangle + \Delta K\right)}{\rho_u(t+\Delta t)\left\langle E(t+\Delta t)\right\rangle_b - \rho_b(t+\Delta t)\left\langle E(t+\Delta t)\right\rangle_u}, \tag{54}$$

where

$$\begin{aligned}
\Delta\left\langle E\right\rangle &\equiv \left\langle E(t+\Delta t)\right\rangle - \left\langle E(t)\right\rangle, \\
\Delta K &\equiv K(t+\Delta t) - K(t).
\end{aligned} \tag{55}$$

Thus, for $\Delta < 0$, the nuclear velocities are rescaled with Eq. (46) and the electronic coefficients with

$$c_N(t+\Delta t) = \begin{cases} c_N^{(0)}(t+\Delta t)\times\left(\pm\sqrt{1+\dfrac{\rho_u(t+\Delta t)}{\rho_b(t+\Delta t)}\beta^2}\right), & N\in N_b \\[3ex] c_N^{(0)}(t+\Delta t)\times\left(\pm\sqrt{1-\beta^2}\right), & N\in N_u \end{cases}. \tag{56}$$

The sign of the rescaling factors is chosen randomly.

## 5. Computational Details

SLED propagates independent nuclear classical trajectories [Eq. (12)] on a mean potential energy surface [Eq. (25)], weighted by electronic populations and coherences computed using the QSD equation with the electronic Hamiltonian set as the localization operator and expanded in the adiabatic energy basis [Eq. (33)]. Total energy conservation is ensured by rescaling velocities (and coefficients when needed) every time step (see Section 4.3). Table 1 summarizes the main features and equations of motion (EOM) in conventional Ehrenfest dynamics and SLED. SM-2 discusses an algorithm for integrating these EOMs, and SM-3 presents Skitten, our novel software implementation of SLED.

Table 1. Summary of the equations of motion (EOM) and features of Ehrenfest dynamics and Ehrenfest dynamics with spontaneous localization (SLED), both in adiabatic representation.

| Features | Ehrenfest Dynamics | SLED |
|---|---|---|
| Quantum EOM | $\dfrac{dA_M}{dt} = -\sum_{N \neq M} e^{-i\gamma_{NM}} \sigma_{MN} A_N$ | $dc_M = -\sum_{N \neq M} e^{-i\gamma_{NM}} \sigma_{MN} c_N dt$ |
| | | $\quad -\dfrac{\kappa}{2}\left(E_M - \langle E \rangle\right)^2 c_M dt + \sqrt{\kappa}\left(E_M - \langle E \rangle\right)c_M dW_M$ |
| | | $\mathbf{A} = \dfrac{\mathbf{c}}{\sum_N |c_N|^2}$ |
| Classical EOM | | $\dfrac{d^2\mathbf{R}_\alpha}{dt^2} = -\dfrac{1}{M_\alpha}\left\langle \Psi \middle| \nabla \hat{H} \middle| \Psi \right\rangle$ |
| | | $\left\langle \Psi \middle| \nabla \hat{H} \middle| \Psi \right\rangle = \sum_N |A_N|^2 \nabla E_N - 2\sum_{N,M>N} \mathrm{Re}\left[ A_M A_N^{*} e^{-i\gamma_{MN}} \right](E_M - E_N)\mathbf{d}_{MN}$ |
| Total energy | Conserved | Rescaling is required for conservation. |
| Decoherence | No decoherence | Decoherence is considered through stochastic processes. |
| Parameters | Parameter free | Localization kernel $\kappa$. |

SLED, Ehrenfest, and surface hopping simulations are based on independent trajectories. Due to the stochastic nature of SLED and surface hoping, each trajectory may have a different set of initial conditions or start from the same initial conditions (realization), but with different random number seeds. In the following, an ensemble of $N_T$ trajectories, which is the product of $N_I$ sets of initial conditions and $N_R$ realizations, is denoted as ($N_I$I, $N_R$R).

We tested SLED with the one-dimensional Tully models, specifically the single avoided crossing (Tully 1), the dual avoided crossing (Tully 2), and the extended coupling with reflection (Tully 3) models.[37] The parameter values used are the same as those proposed by Tully. All simulations started from the lower state (State 1). Initial conditions were sampled from a 1-D Wigner distribution for a Gaussian wave packet (GWP)[51]

$$P_{GWP}(x,p) = \frac{1}{\pi\hbar} e^{-\frac{(x-x_0)^2}{2\Delta x^2}} e^{-\frac{(p-p_0)^2}{2\Delta p^2}}, \tag{57}$$

where $x_0$ and $p_0$ are the position and momentum centers, $\Delta x$ is the spatial width, and $\Delta p = \hbar / (2\Delta x)$ is the momentum width. $x_0$ = −10 a.u. for Tully 1 and −15 a.u. for Tully 2 and Tully 3. Two sets were computed of each model, one with low initial momentum, $p_0$ = 10 a.u., and another with high initial momentum, $p_0$ = 30 a.u. For Tully 2, extra sets of dynamics with momentum varying from 10 to 50 a.u. were also used to compute the lower transmission (State 1 population of trajectories in the positive $x$ coordinate) as a function of initial momentum. Following Tully,[37] the spatial width was $\Delta x = 20 / p_0$.

We also tested SLED on multidimensional models. In this case, we used the spin-boson Hamiltonian model[52] in adiabatic representation[53] coupled to $N$ oscillators (ASBH-$N$). We run simulations with ASBH-10, a model we proposed in Ref. [54] to test nonadiabatic dynamics in a regime of tens of picoseconds. It has also been used for benchmark calculations in Ref. [55] We also employed a 5-dimensional model (SBH-5 (diabatic) and ASBH-5 (adiabatic)), designed for dynamics in the sub-picosecond regime with strong initial coherence. Both models are characterized in SM-4.

Fewest-switches surface hopping (FSSH) and decoherence-corrected fewest-switches surface hopping (DC-FSSH) calculations were performed, too. Decoherence corrections were considered using the simplified-decay of mixing algorithm, adopting the standard value of 0.1 a.u. parameter.[12] Momentum was maintained after frustrated hopping attempts. For the lower transmission calculation in Tully 2, initial conditions were generated as discussed above, with $x_0 = -15$ a.u., and $p_0$ varying from 10 to 50 a.u with a step of 1 a.u. The total propagation time ranges from 150 fs to 30 fs, depending on $p_0$, with a timestep of 0.1 fs for nuclear integration and 0.005 fs for electronic coefficient integration.

We performed quantum wave packet dynamics simulations on three one-dimensional Tully models. Details are given in SM-5. In brief, the wave function was propagated numerically exactly using the "exact" keyword as implemented in the MCTDH software.[56] In each quantum dynamics simulation, we placed a Gaussian wave packet at the lower diabatic state, with its position centered at $-20$ a.u. and an initial momentum of either 10 a.u. or 30 a.u. to build the initial wave function. The width of the wave packet was set to $\Delta x = 20/p_0$, where $p_0$ denotes the initial momentum. The mass was set to 2000 a.u. For the transmission calculation in Tully 2, the initial momentum was varied from 10 to 50 a.u. with a step size of 2.5 a.u., with additional points included in regions with significant curvature. After simulating the wave packet dynamics in the diabatic representation, the adiabatic populations and the adiabatic coherence elements are calculated using the *adpop* analysis program inside the MCTDH package.

Besides FSSH, DC-FSSH, and quantum dynamics, we also computed the dynamics of the Tully models using *ab initio* multiple spawning (AIMS).[35, 57] All those procedures and results are described in SM-6.

We performed quantum wave packet dynamics on the SBH-5 model using the multi-configurational time-dependent Hartree (MCTDH) expansion. Details on these calculations can be found in SM-7. The initial wave packet was placed at the minima of the second diabatic state with an initial momentum of $p_k = \sqrt{\hbar M_k \omega_k}$, where $\omega_k$ and $M_k$ are the frequency and mass of the $k^{th}$ mode. The wave packet is launched from the first diabatic state and propagated up to 1000 fs with an integration time step of 0.1 fs. We have conducted systematic convergence tests by varying the number of single-particle functions (SPFs), $n_{SPF}$, and the grid size of the primitive basis functions, $N_{GS}$. As shown in SM-7, the simulations achieved reasonable convergence by systematically increasing the basis size. Adiabatic populations were computed to serve as reference data in Section 6.2.

For completeness, we also computed FSSH and DC-FSSH results for the 5-D model. In this case, 5000 trajectories were propagated with a time step of 0.01 fs for the classical coordinates, up to 1000 fs. Initial conditions were sampled as explained in Ref. [55] Extra details are presented in SM-8.

Initial conditions for SLED, Ehrenfest, surface hopping, and AIMS were generated using the Newton-X CS program.[58] SLED and Ehrenfest dynamics were done with Skitten. Surface hopping simulations were propagated using Newton-X NS (3.5.2).[58] Exact quantum dynamics and MCTDH propagations were performed using the Heidelberg MCTDH package[56] version 8.6.2. AIMS was done with Legion.[19]

# 6. Results

## 6.1. Test Cases with 0- and 1-Dimensional Models

A significant feature of the QSD equation is that, upon spontaneous localization, the ensemble is distributed according to the Born rule. To illustrate this point, we applied SLED to the simple harmonic oscillator, a model for which the Born rule prediction is precisely known. We prepared the oscillator in a quantum superposition of the ground and the first two excited states:

$$\left|\psi\left(0\right)\right\rangle = \sqrt{\frac{1}{6}}\left|1\right\rangle + \sqrt{\frac{2}{3}}\left|2\right\rangle + \sqrt{\frac{1}{6}}\left|3\right\rangle. \tag{58}$$

The probability of measuring state $\left|1\right\rangle$ or $\left|3\right\rangle$ is 1/6, while for state $\left|2\right\rangle$, it is 2/3. After spontaneous localization, these exact probabilities must be observed. As illustrated in Figure 1, this is precisely what happens when we solve Eq. (33) for this model. Naturally, this model has no nuclear degrees of freedom (0-D), and $\sigma_{MN}$ is zero.

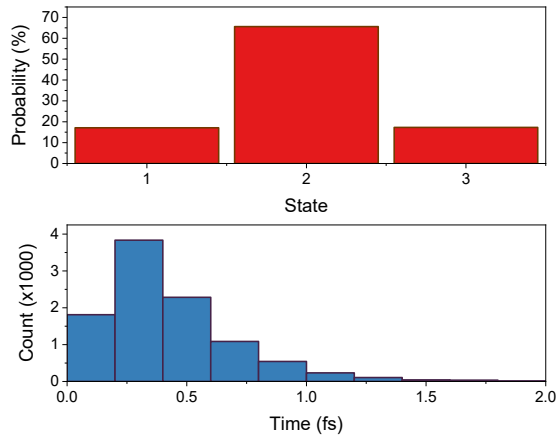

Figure 1. Spontaneous localization in a one-dimensional harmonic oscillator initially in the coherent state [Eq. (58)], involving the ground and the two first excited states. (Top) Distribution of localized states at the end of dynamics. As expected, the results match those from the Born rule, with probabilities of 1/6, 2/3, and 1/6 for states 1, 2, and 3, respectively. (Bottom) Distribution of localization time. Simulations with 10,000 realizations with $\kappa$ = 0.25 a.u.

Moving to slightly more realistic models, we tested SLED on the three one-dimensional (1-D) Tully models with standard parameters.[37]

Figure 2 shows how SLED localizes the trajectories. The figure displays results for the Tully 2 model; however, the behavior is similar to that of the other models as well. The top and middle panels show trajectories started from the same initial conditions. In both cases, the system energy coincides with state 1 energy until the system reaches the avoided crossing, around 20 fs. There, coherence between states 1 and 2 builds, and the system's energy evolves as an average value, similar to the Ehrenfest dynamics.

Nevertheless, SLED and Ehrenfest results diverge after 30 fs. In Figure 2-top, the SLED trajectory stochastically localizes in state 2, whereas in Figure 2-middle, it localizes in state 1. The Ehrenfest result, as expected, does not localize and persists in a state superposition until the end of the simulation. This example uses $\kappa$ = 0.30 a.u., chosen to minimize the deviation between the SLED and exact results for both population and coherence (Figure 2-bottom).

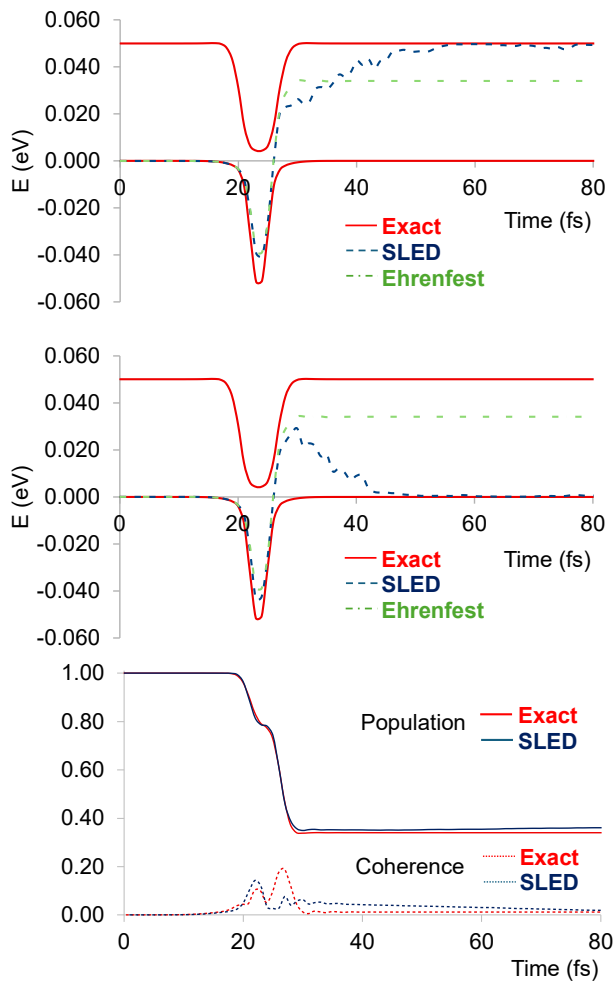

Figure 2. Tully 2 model simulations. The top and middle graphs display the potential energy of individual trajectories as a function of time, starting from the same initial conditions. In each case, the mean energy of the SLED dynamics localizes either in state 2 (top) or 1 (middle). The standard Ehrenfest result is shown for comparison. Mean values of state 1 population (solid line) and coherence between states 1 and 2 (dashed line), averaged over 10,000 trajectories (100I, 100R), are shown in the bottom graph. This graph also shows the exact result. SLED was computed with $\kappa = 0.30$ a.u. The initial momentum was $p_0 = 30$ a.u.

Populations in Figure 2-bottom are computed with Eq. (9). The coherence is the module of the term defined in Eq. (10). MQCD coherences neglect the phase factor $\exp(i\gamma_{KL})$ appearing in the state definition, Eq. (18), because these phases are random between independent trajectories. Thus, if they were accounted for in the average over the trajectories, the mean coherence would tend to zero.

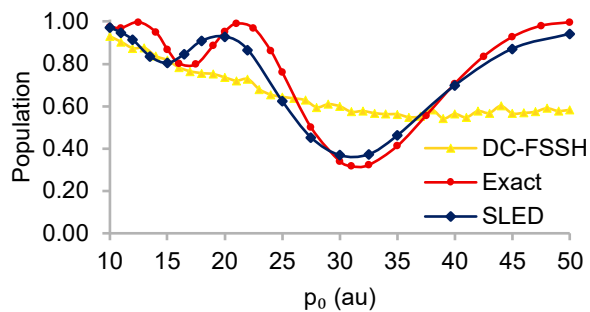

Figure 3. Asymptotic state 1 population of Tully 2 as a function of the initial momentum computed with SLED, DC-FSSH, and quantum wave packet propagation (exact). SLED: $\kappa$ = 0.3 a.u. with 10,000 trajectories (100I, 100R) for each $p_0$. DC-FSSH: 2000 trajectories (2000I, 1R) for each $p_0$.

Mean populations for initial momentum values between 10 and 50 a.u. are shown in Figure 3. We adopted $\kappa$ = 0.3 a.u., which yielded satisfactory results for the entire momentum range. At this point, our goal is to demonstrate how SLED can yield qualitatively correct results where FSSH, corrected for decoherence using the widespread simplified decay of mixing, fails, a result that has also been previously reported in Ref. [59]

A central aspect of SLED is that it is developed to provide a balanced description of both populations and coherences. We illustrate this point in Figure 4, where SLED is employed for all three Tully models (using the standard parameterization) with two different initial momentum values and compared with the exact result. In each case, we chose $\kappa$ so that it gave satisfactory results for both quantities. This strategy is relevant because it is possible to optimize, for instance, the population at the expense of coherence.

We observe in Figure 4 that across the entire range of models and momenta, we can determine $\kappa$ values that describe the system, at least semi-quantitatively. Overall, populations are well represented in almost all cases. Coherences, however, show slight but revealing differences from the benchmarks. Particularly, for Tully 1 with low momentum (Figure 4, top left), we were unable to find a $\kappa$ value that simultaneously yielded accurate values for population and coherence. We opted for displaying the result for $\kappa$ = 10 a.u., which correctly predicts the decoherence time, but underestimates the population. We could have, alternatively, adopted $\kappa$ = 0.3 a.u. In this case, the population would perfectly match the exact value, but the decoherence time would be excessively long.

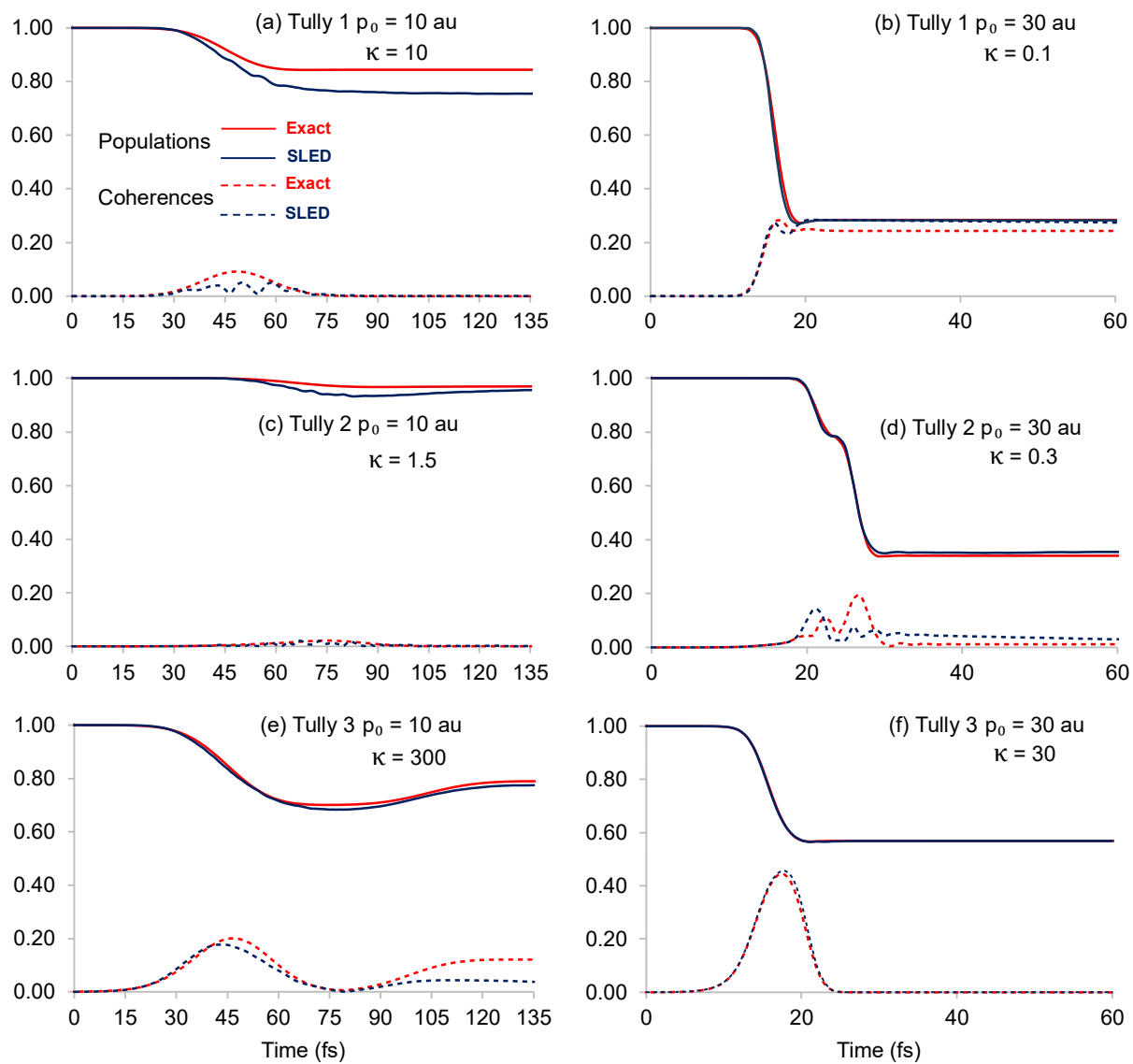

Figure 4. State 1 population (solid lines) and coherence (dashed lines) between states 1 and 2 for the Tully models 1 ((a) and (b)), 2 ((c) and (d)), and 3 ((e) and (f)). The graphs on the left have a central initial momentum of 10 a.u., and those on the right have a central initial momentum of 30 a.u. Each simulation contains 10,000 trajectories (100I, 100R).

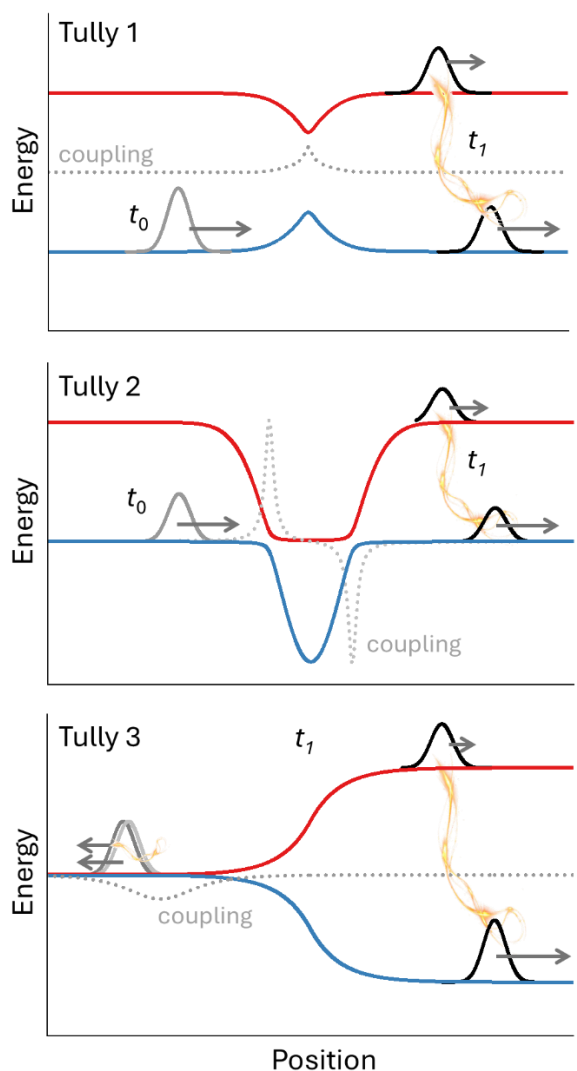

Figure 5. Schematic illustration of the wavepacket propagation in the three Tully models. Each figure illustrates the two adiabatic states (blue and red) and the nonadiabatic coupling (light gray) as a function of the position. In Tully 1 and 2, $t_0$ represents the system at an earlier time before it reaches the coupling region, while $t_1$ shows it after it has left the coupling region. Tully 3 only shows the system at the later time $t_1$, with transmitted (right) and reflected (left) components. The yellow curves represent the electronic coherences.

Tully 3 with low momentum (Figure 4-bottom-left) also deserves attention. The exact coherence peaks at 45 fs and then again at 105 fs, after which it remains constant. The earlier peak corresponds to the rapid decoherence of the transmitted wave packet. In contrast, the second peak arises from the overlapping reflected wavepackets, which move with similar speeds due to the nearly degenerate states in the negative region of the potential (Figure 5-bottom). SLED perfectly describes the earlier coherence peak in Figure 4, bottom left. However, it cannot build sufficient coherence to capture the later peak.

These limitations in SLED description arise from the use of a constant localization kernel $\kappa$, which, as discussed in Section 7, is inadequate for complex coherence dynamics and should be replaced by a time- and phase-space-dependent function.

To support SLED testing, we benchmarked the Tully models against established methods, including FSSH, DC-FSSH, Ehrenfest dynamics, AIMS, and quantum dynamics propagation. These results are shown in SM-6. Analogous benchmark results for the Tully models with other methods are reported in Refs. [60, 61]

For most methods, population is described fairly well within ±5% of the reference value (quantum dynamics). The exception is Tully2, with an initial momentum of 30 a.u., which exhibits significant dispersion in the results. Notably, the only methods that agree with the reference are FSSH and Ehrenfest. Concerning coherence, the results are uneven. Unsurprisingly, FSSH and Ehrenfest miss decoherence while AIMS nicely matches the reference coherences for most tests, including the challenging double peak in Tully 3 with low momentum.

Notably, Ehrenfest dynamics simulation perfectly matches the reference population in Tully1 with an initial momentum of 10 a.u., a case where SLED did not perform well (Figure 4, top left). However, the Ehrenfest simulation is right for the wrong reason. Its coherence persists in regions where the exact coherence vanishes. Hence, its correct population stems from an unphysical population exchange.

A revealing case is Tully 1 with $p_0$ = 30 a.u., which exhibits long-lived coherence. (We have already seen this feature in Figure 4-top-right). This unusual situation arises from the wavepacket thawing in the flat asymptotic states, compensating for the dephasing between the transmitted components in the low and high states. Now, FSSH's lack of decoherence plays in its favor, yielding the best match with the reference. The *ad hoc* decoherence in DC-FSSH is entirely incorrect, and AIMS also fails, predicting too rapid decoherence.

AIMS failure in this case suggests that its frozen Gaussian bases may not accurately describe the thawing process. In this model, the trajectory passes through the coupling region only once, so there is only a single Gaussian in each state, and a single frozen Gaussian cannot describe the larger phase space occupied by the thawed wavefunction. Methods to improve this description using multiple frozen Gaussians have been discussed.[62]

## 6.2. Test Cases with Multidimensional Models

We can illustrate the potential of SLED through multidimensional nonadiabatic dynamics, employing the spin-Boson Hamiltonian model with N dimensions (SBH-N) in an adiabatic representation (ASBH-*N*). We worked with two models, one with 5 dimensions and the other with 10 dimensions (see Section 5).

Figure 6 shows the energy evolution in a 100-fs extract of two trajectories starting from the same initial conditions propagated with SLED and Ehrenfest dynamics for ASBH-5. In both cases, the current (mean) energy tends to be closer to the lower than to the higher state. Nevertheless, as expected, SLED tends to be more localized than Ehrenfest, with the current energy closer to the eigenstate than Ehrenfest is. Note that after 420 fs, SLED energy delocalizes, indicating that coherence is rebuilt. This is a direct result of vibrational motion, which causes the molecule to return to coupling regions.

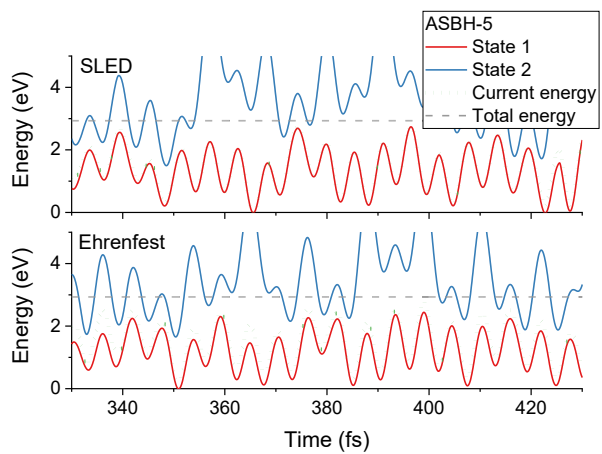

Figure 6. Energy as a function of time for a single trajectory with the ASBH-5 model. Top: SLED ($\kappa$ = 0.04 a.u.). Bottom: Ehrenfest dynamics starting from the same initial conditions.

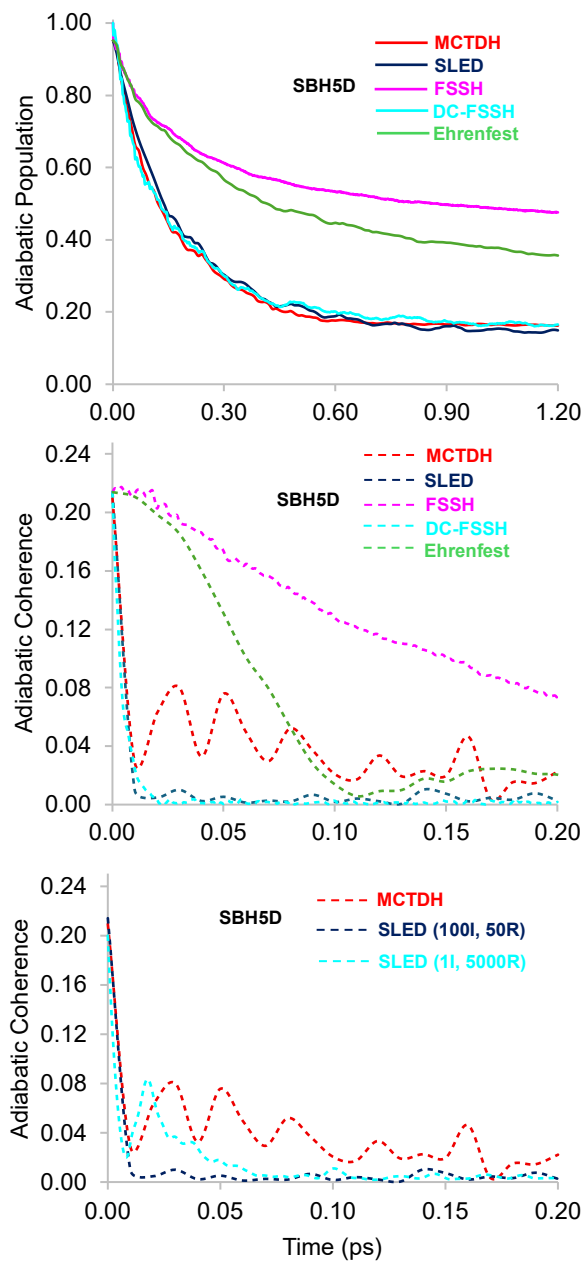

Figure 7. Adiabatic populations and coherence in ASBH-5. Top: Population of state 2 as a function of time with SLED (100I, 50R), MCTDH ($n_{SPF}$ = 12, $N_{GS}$ = 24; see SM-7), FSSH (5000I, 1R), DC-FSSH (5000I, 1R), and Ehrenfest dynamics (5000I, 1R). Middle: Coherence as a function of time with the same methods as in the Top graph. Bottom: Coherence as a function of time with SLED (100I, 50R), SLED (1I, 5000R), and MCTDH ($n_{SPF}$ = 12, $N_{GS}$ = 24). All SLED calculations with $\kappa$ = 0.04 a.u.

By construction, SBH-5 has strong nonadiabatic coupling near the ground state minimum. As a consequence, it exhibits quantum coherence between states 1 and 2 from the outset, at time zero. For this reason, the initial population is not the unity, reflecting a 4.8:95.2 ratio between states 1 and 2. With MCTDH, Ehrenfest, and SLED, it is straightforward to model such an atypical initial condition. Ehrenfest and SLED were initiated in the superposition $\left|\Psi(0)\right\rangle = 0.219\left|1\right\rangle + 0.976\left|2\right\rangle$. Surface hopping, however, is not suited for that. In our FSSH and DC-FSSH simulations, we launched all trajectories from state 2, but with the same electronic state superposition, which is not an entirely consistent setup.

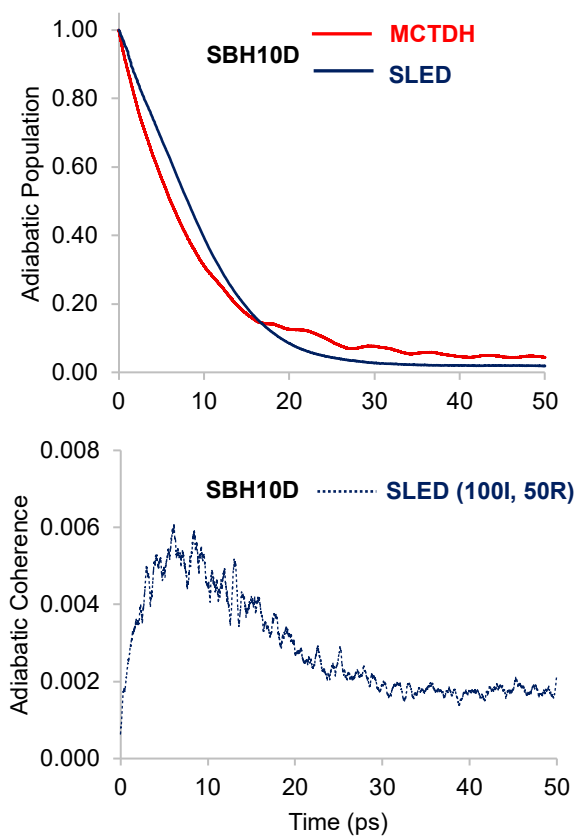

Figure 8. Adiabatic populations and coherence in ASBH-10. Top: Populations as a function of time with SLED (100I, 50R) and MCTDH ($n_{SPF}$ = 12, $N_{GS}$ = 64) from Ref. [55] Bottom: Coherence with SLED (100I, 50R). All SLED calculations with $\kappa$ = 4.0×10⁻⁴ a.u.

The mean population of State 2 as a function of time is shown in Figure 7-top for ASBH-5 computed with SLED and a few other methods. The reference result is MCTDH. In the case of SLED, $\kappa$ was chosen to give the best match with MCTDH. Ehrenfest and FSSH underestimate the population transfer, resulting in too long lifetimes. SLED and DC-FSSH perfectly match with the MCTDH population, although in the case of SLED, this is by construction, due to the carefully tuned $\kappa$.

The middle panel of Figure 7 shows the adiabatic coherence evolution for the same methods. Compared to MCTDH, SLED accurately describes the initial decoherence that occurs within 0.01 ps; however, it fails to capture the coherence revival observed in MCTDH. The same behavior is observed in DC-FSSH. Unsurprisingly, Ehrenfest dynamics delivers a completely wrong coherence behavior. What is somewhat unexpected is that FSSH coherence evolution is even worse than that of the Ehrenfest dynamics.

The reason SLED misses the coherence revival stems from the independent trajectory approach. The revival reflects vibronic coherences that cannot be accounted for within classical trajectories. To illustrate this feature, we again plot the coherence as a function of time in Figure 7-bottom with SLED and MCTDH. Additionally, we included the coherence obtained with SLED (1I, 10,000R), which represents 10,000 realizations of the same initial conditions. The strong initial correlation between the realizations in this setup is enough to create the revival.

Figure 8 shows the results for ASBH-10. Once again, the SLED population matches the MCTDH reference, after tuning $\kappa$. The coherence is shown in Figure 8-bottom, but unfortunately, the MCTDH adiabatization procedure was too computationally costly, and we do not have the wavepacket coherence to compare with. The minimal coherence (<0.01) during the entire dynamics helps rationalize the long lifetime of this model.

## 7. Discussion

The results in the previous section demonstrate the potential of SLED as a framework for accurate nonadiabatic dynamics in the limit of the independent-trajectories MQCD approximation. Specifically, SLED aims to provide a balanced treatment of diagonal (populations) and off-diagonal (coherences) terms of the electronic reduced density matrix (RDM) in the energy basis. This feature distinguishes it from other MQCD methods, which have focused on population description, adopting an *ad hoc* treatment of coherences.

Commonly, these *ad hoc* treatments, either via direct modification of the electronic equations of motion[11, 27, 29, 30, 39] or post-processing of the electronic coefficients,[7, 12, 28, 40, 41] introduce non-linearities in the propagation of the electronic RDM that may have detrimental effects on the simulations, especially allowing superluminal signaling. In turn, SLED is built upon the Gisin-Percival quantum-state diffusion (QSD) equation,[15] which ensures that any non-linearity introduced at the trajectory level is fully compensated by control terms that lead to linear dynamics of the density at the trajectory ensemble level.

The reliance of SLED on the Gisin-Percival QSD equation is beneficial because it also ensures that the Born rule is observed and that the evolution of stochastic trajectories corresponds to a Lindblad propagation of the electronic RDM. This connection is relevant because it implies that SLED dynamics inherits other crucial features of the Lindblad master equation: linearity, trace conservation, and complete positivity (linear CPTP map).[20] Moreover, the SLED connection to the Lindblad equation creates an underexplored bridge between two research fields that have evolved independently for too long: MQCD and master equation dynamics.[25]

As we have seen in the examples discussed in Section 6, in the absence of nonadiabatic couplings, SLED drives the electron-nuclear molecular dynamics toward classical states in the adiabatic electronic energy basis. This is achieved not by directly propagating the electronic RDM toward a diagonal matrix, but through an ensemble of stochastic trajectories, each ending in a pure state, localized on a specific energy eigenvalue. The ensemble of such pure states statistically yields a diagonal electronic RDM.

Despite the satisfactory quality of the results we discussed, SLED is not yet ready for general dynamics. In the test cases, we either used generic values for the electron-nuclear coupling parameter $\kappa$ or adjusted it to match a reference result. Naturally, that is not a desirable feature in a ready-for-production MQCD theory.

The next step in the SLED development must focus on developing a theory for the localization kernel $\kappa$ enabling it to be estimated from the system features. We know, for instance, that decoherence should depend on the topographic differences between the potential energy surfaces.[5] Such information is not contained in the SLED equations of motion, and should be transferred through $\kappa$.

We can also expect situations where the system moves from regions with such parallel surface topography to areas with significant gradient differences. Once more, SLED should rely on $\kappa$ to learn that. Thus, $\kappa$ should not be constant, but a function of time, $\kappa(t)$. Although all theoretical discussion in Section 4.2 didn't make reference to this time dependence, we anticipate that the derived equations of motion remain valid and with the desired properties, as long as $\kappa(t)$ remains *time-local*, meaning it depends only on the current time and no other. This assessment is a direct consequence of the work by Watanabe *et al.*[63] who showed that complete positivity is ensured for time-local jumping operators in the Lindblad equation. (The other properties, linearity and trace conservation, are trivially ensured.)

Including time dependence on $\kappa$ may not be sufficient for an adequate description of the dynamics. In Tully 3, for instance, we may expect that the decoherence time between transmitted wavepackets in the two states is much shorter than the decoherence time for a wavepacket reflected. Therefore, the localization kernel $\kappa$ should, in this case, depend on nuclear position. Moreover, we have observed that the optimal $\kappa$ value differs between high and low momentum (Figure 4), meaning that it should also depend on this variable. Thus, we may expect that a general expression for $\kappa$ should depend on time and nuclear phase space variables. Such a generalization still ensures that a linear CPTP map is possible within the framework of the fully consistent classical-quantum (FCQC) dynamics proposed by Oppenheim and co-workers.[64-66]

So far, $\kappa$ has been assumed to contain information about the topographic differences between the two states. Naturally, SLED must be valid for an arbitrary number of states, which requires generalizing $\kappa$ to account for information from all pairs of states. The most straightforward way to achieve it is to modify the localization operator in Eq. (29) to

$$\sum_{n} \hat{L}_n \rightarrow \hat{\Lambda}\hat{H}, \tag{59}$$

where $\hat{\Lambda}$ is a self-adjoint operator commuting with $\hat{H}$, defined in terms of the energy eigenstates as

$$\hat{\Lambda}\big|N(t)\big\rangle = \sqrt{\kappa_N(t;\mathbf{z})}\big|N(t)\big\rangle, \tag{60}$$

where $\kappa_{MN}(t;\mathbf{z})$ is a positive function of time with $\kappa_{MN}(t;\mathbf{z}) = \kappa_{NM}(t;\mathbf{z})$. $\mathbf{z}$ represents the nuclear phase space coordinates, on which $\kappa$ depends parametrically. In this way, each adiabatic state has its own electron-nuclear coupling parameter, conveying information from all relevant state pairs.

The electronic equation of motion in SLED, Eq. (33), is trivially modified to accommodate the state- and time-dependent functions $\kappa_{MN}(t)$ from Eq. (60). Dependence on $\mathbf{z}$, however, requires more extensive modifications toward FCQC dynamics. Either way, we must still explicitly define these functions. We are currently working on this non-trivial task. However, because it will require extensive new theoretical developments, we reserve this discussion for future work, restricting the present paper to a proof-of-principle presentation of SLED and of its implementation in Skitten software.

Finally, supposing that we succeed in deriving a localization kernel $\kappa_{MN}(t;\mathbf{z})$ with all those functional dependencies, there is still one last challenge to face: the failure of the independent trajectory approximation to deal with strong vibronic coherences, as we saw in the ASBH-5 example. Far from being a problem of SLED only,[67] we must consider that the MQCD description of advanced coherent problems will require interacting trajectories; a target that has been under the radar of diverse groups for a while already.[60, 61, 68-70] Nonetheless, if such a development effort must be undertaken, it may be beneficial to focus it on a well-grounded theoretical method, such as SLED.

## 8.  Conclusion

We present SLED, a novel mixed quantum-classical dynamics method that incorporates decoherence through spontaneous localization, governed by the Gisin-Percival quantum-state diffusion equation. Built upon Ehrenfest dynamics, SLED retains the independent trajectory approximation while restoring a physically grounded treatment of coherence loss. Through a series of test cases, ranging from 0-D harmonic models to multidimensional spin-boson Hamiltonians, we demonstrated that SLED yields accurate populations and semi-quantitative coherences, often surpassing the performance of conventional decoherence-corrected methods.

Crucially, SLED inherits the formal structure of the Lindblad equation. As such, the ensemble electronic dynamics form a linear, trace-preserving, and completely positive (CPTP) map, ensuring physically consistent nonadiabatic evolution derived from first principles. This connection bridges mixed quantum-classical dynamics with open quantum system theory, establishing a foundation for further developments, including flexible pointer states.

Despite its promising results for the test cases we explored here, SLED remains a proof of principle. Its current implementation depends on open-valued localization kernel controlling decoherence strength. Our tests demonstrated that this quantity is not constant in general. Future work will aim to derive the localization kernel from system features and generalize it to a time- and phase-space-dependent function. We anticipate that this development will further consolidate SLED as a robust and theoretically consistent framework for nonadiabatic dynamics simulations.

## Supplementary Material

The supplementary material contains: 1. Relation between QSD and Lindblad equations; 2. An algorithm to integrate the SLED EOMs; 3. Details on Skitten, the software implementation to simulate SLED; 4. Description of the SBH-5 and SBH-10 models; 5. Information on the quantum dynamics propagation of 1-D models; 6. Benchmark of results for Tully models with established methods; 7. Details on the MCTDH propagation of SBH-5; 8. Additional DC-FSSH results for ASBH-5.

## Acknowledgments

This work received support from the French government under the France 2030 investment plan as part of the Initiative d'Excellence d'Aix-Marseille Université (A*MIDEX AMX-22-REAB-173 and AMX-22-IN1-48) and from the European Research Council (ERC) Advanced Grant SubNano (grant agreement 832237). The authors acknowledge the Centre de Calcul Intensif d'Aix-Marseille for granting access to its high-performance computing resources. SM acknowledges the Wroclaw Networking and Supercomputing Center for generously providing computational resources used for performing several of the simulations presented in this study. He also thanks Prof. H. D. Meyer for access to the *adpop* program.

## Author contributions

Conceptualization: MB; Formal analysis: MB; Funding acquisition: MB; Investigation: AAT, RSM, SM; Methodology: MB; Project administration: MB; Software: MB; Supervision: MB; Validation: SM; Visualization: AAT, RSM; Writing – original draft: MB; Writing – review & editing: AAT, RSM, SM, MB.

**Supplementary Material**

# Ehrenfest Dynamics with Spontaneous Localization

Anderson A. Tomaz, Rafael S. Mattos, Saikat Mukherjee, Mario Barbatti

## Table of contents

## SM-1. Relation Between the QSD and Lindblad Equations

The Gisin-Percival QSD equation is derived in different ways in Refs. [1, 2] Here, we follow Gisin and Percival's discussion[3] to show that the QSD equation

$$d|\Phi\rangle = -\frac{i}{\hbar}\hat{H}|\Phi\rangle dt - \frac{1}{2}\left(\hat{L}^\dagger - \left\langle L^\dagger\right\rangle\right)\left(\hat{L} - \left\langle L\right\rangle\right)|\Phi\rangle dt + \left(\hat{L} - \left\langle L\right\rangle\right)|\Phi\rangle dW \tag{1}$$

implies an ensemble density $\bar{\rho}$ that evolves with the Lindblad equation

$$\frac{d\bar{\rho}}{dt} = -\frac{i}{\hbar}\left[\hat{H},\bar{\rho}\right] + \hat{L}\bar{\rho}\hat{L}^\dagger - \frac{1}{2}\left\{\hat{L}^\dagger\hat{L},\bar{\rho}\right\}. \tag{2}$$

To see that, we start with the Itô product rule,

$$d\left(|\Phi\rangle\langle\Phi|\right) = d\left(|\Phi\rangle\right)\langle\Phi| + |\Phi\rangle d\left(\langle\Phi|\right) + d|\Phi\rangle d\langle\Phi|, \tag{3}$$

Which, after some algebraic manipulation of Eq. (1), gives

$$d\rho = -\frac{i}{\hbar}\left[\hat{H},\rho\right]dt + \left(\hat{L}\rho\hat{L}^\dagger - \frac{1}{2}\left\{\hat{L}^\dagger\hat{L},\rho\right\}\right)dt + \left(\left(\hat{L} - \left\langle L\right\rangle\right)\rho + \rho\left(\hat{L}^\dagger - \left\langle L^\dagger\right\rangle\right)\right)dW, \tag{4}$$

where $\rho = |\Phi\rangle\langle\Phi|$, and we assumed $dt^2 = 0$ and $dtdW = 0$. $[\bullet]$ represents the commutator and $\{\bullet\}$ the anticommutator.

Given that $\langle dW\rangle = 0$, the density $\bar{\rho}$ of an ensemble of realizations evolves as Eq. (2). This derivation can be trivially extended to more general $\sum_n \hat{L}_n$ cases.

## SM-2. Solving the Coupled Equations

This section presents our algorithmic implementation of integrating the equations of motion. We give it for completeness, but stress that it is not part of SLED. Other (and more efficient) integrators could be used.

At time *t*, we employ the velocity Verlet algorithm to update the position of a nucleus $\alpha$ at the full step $\Delta t$ through

$$\mathbf{R}_\alpha\left(t + \Delta t\right) = \mathbf{R}_\alpha\left(t\right) + \mathbf{v}_\alpha\left(t\right)\Delta t + \frac{1}{2}\mathbf{a}_\alpha\left(t\right)\Delta t^2, \tag{5}$$

and the velocity at half-step with

$$\mathbf{v}_\alpha\left(t + \frac{\Delta t}{2}\right) = \mathbf{v}_\alpha\left(t\right) + \frac{1}{2}\mathbf{a}_\alpha\left(t\right)\Delta t, \tag{6}$$

where

$$\mathbf{a}_\alpha(t) =$$

$$-\frac{1}{M_\alpha}\left[\sum_N \left|A_N(t)\right|^2 \nabla_\alpha E_N(t) - 2\sum_{N,M>N} \mathrm{Re}\left[A_M(t)A_N(t)^* e^{-i\gamma_{MN}(t)}\right](E_M(t)-E_N(t))\mathbf{d}_{MN}(t)\right]. \quad (7)$$

Energies, gradients, and nonadiabatic coupling vectors are then computed for this new geometry.

Next, we update the wavefunction coefficients $\mathbf{A} = \mathbf{c}/||\mathbf{c}||^2$ in one time step by integrating $\mathbf{c}$ with 4th-order Runge-Kutta

$$\mathbf{c}(t+\Delta t) = \mathbf{c}(t) + \frac{1}{6}\left(\mathbf{k}_1 + 2\mathbf{k}_2 + 2\mathbf{k}_3 + \mathbf{k}_4\right), \quad (8)$$

where

$$\begin{aligned}
\mathbf{k}_1 &= \mathbf{f}\left(t, \mathbf{c}(t)\right), \\
\mathbf{k}_2 &= \mathbf{f}\left(t+\frac{\Delta t}{2}, \mathbf{c}(t)+\frac{\Delta t}{2}\mathbf{k}_1\right), \\
\mathbf{k}_3 &= \mathbf{f}\left(t+\frac{\Delta t}{2}, \mathbf{c}(t)+\frac{\Delta t}{2}\mathbf{k}_2\right), \\
\mathbf{k}_4 &= \mathbf{f}\left(t+\Delta t, \mathbf{c}(t)+\Delta t\mathbf{k}_3\right).
\end{aligned} \quad (9)$$

In these equations,

$$f_M(t,\mathbf{c}) \equiv -\sum_{N\neq M} e^{-i\gamma_{NM}} \sigma_{MN} c_N - \frac{\kappa}{2}\left(E_M - \langle E\rangle\right)^2 c_M + \sqrt{\kappa}\left(E_M - \langle E\rangle\right)\frac{dW_M}{dt}c_M, \quad (10)$$

and

$$\sigma_{MN} = \mathbf{v}\cdot\mathbf{d}_{MN}. \quad (11)$$

The complex stochastic process is modeled as

$$\frac{dW_M}{dt} = \frac{\mathcal{N}_1(0,1)}{\sqrt{\Delta t}} + i\frac{\mathcal{N}_2(0,1)}{\sqrt{\Delta t}}, \quad (12)$$

where $\mathcal{N}_1(0,1)$ and $\mathcal{N}_2(0,1)$ are independent random variables drawn from a normal distribution with a mean of 0 and a variance of 1. Note that different random values are sampled for each $M$, but the same values are used for all $\mathbf{k}_i$ in the time step.

Eqs. (9) require $E_N$ and $\mathbf{d}_{NM}$ (to get $\sigma_{NM}$) computed at the half step for geometry

$$\mathbf{R}_\alpha\left(t+\frac{\Delta t}{2}\right)=\mathbf{R}_\alpha\left(t\right)+\frac{1}{2}\mathbf{v}_\alpha\left(t\right)\Delta t+\frac{1}{8}\mathbf{a}_\alpha\left(t\right)\Delta t^2. \tag{13}$$

Alternatively, they can be obtained via a mid-point interpolation

$$E_N\left(t+\frac{\Delta t}{2}\right)\approx\frac{1}{2}\left(E_N\left(t\right)+E_N\left(t+\Delta t\right)\right),$$
$$\mathbf{d}_{NM}\left(t+\frac{\Delta t}{2}\right)\approx\frac{1}{2}\left(\mathbf{d}_{NM}\left(t\right)+\mathbf{d}_{NM}\left(t+\Delta t\right)\right). \tag{14}$$

Eqs. (9) also require $\mathbf{v}$ (to get $\sigma_{NM}$) to be computed at the half step and $\mathbf{c}$ computed at the half and full steps. They are approximated self-consistently. In the first iteration, they are approximated for their value at $t$:

$$\mathbf{v}_\alpha\left(t+\Delta t\right)\approx\mathbf{v}_\alpha\left(t+\frac{\Delta t}{2}\right)+\frac{1}{2}\mathbf{a}_\alpha\left(t\right)\Delta t. \tag{15}$$

After $\mathbf{c}$ is updated with Eq. (8), the acceleration at $t+\Delta t$ is computed with (7), and velocity Verlet is once more invoked to update the velocity

$$\mathbf{v}_\alpha\left(t+\Delta t\right)=\mathbf{v}_\alpha\left(t+\frac{\Delta t}{2}\right)+\frac{1}{2}\mathbf{a}_\alpha\left(t+\Delta t\right)\Delta t. \tag{16}$$

Finally, the second-iteration values of $\mathbf{c}(t+\Delta t)$ and $\mathbf{v}(t+\Delta t)$ are obtained with Eqs. (8) and (16).

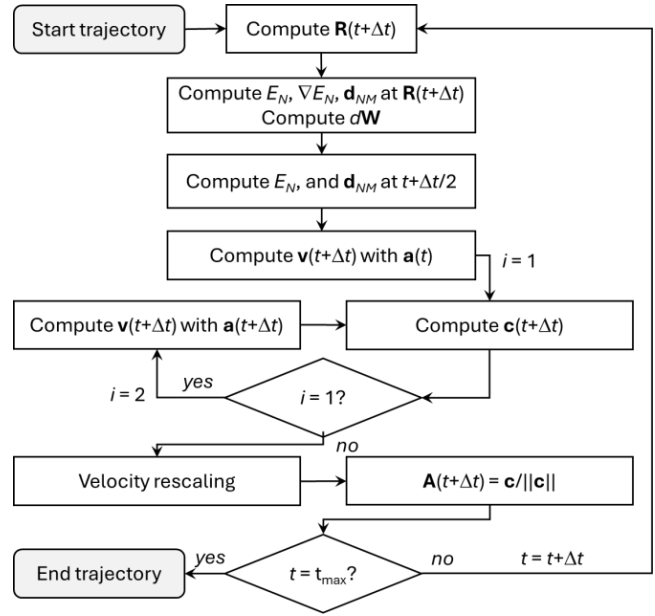

Figure 1. Coupled quantum-classical EOM integration using velocity Verlet and 4th-order Runge-Kutta.

At this point, we make $\mathbf{v}_\alpha(t+\Delta t) = \mathbf{v}_\alpha^{(0)}(t+\Delta t)$ and apply the velocity rescaling algorithm to impose total energy conservation, as discussed in the main text.

This algorithm, which combines the symplectic advantages of the velocity Verlet method with the accuracy of the 4th-order Runge-Kutta method, is illustrated in Figure 1.

## SM-3. Software Implementation

We implemented SLED in a Fortran program called Skitten, Stochastic Schrödinger Cats. The program integrates the equations of motion with the methods described in 0, ensuring total energy conservation for SLED through the rescaling discussed in the main text. Conventional Ehrenfest can be trivially obtained by setting $\kappa = 0$.

Several analytical models are available in Skitten, including all three Tully models,[4] the 1D double-arch model,[5] and the adiabatic spin boson Hamiltonian (SBH) in arbitrary $N$ dimensions.[6] The program is also highly modular, making it straightforward to add additional models. Skitten delivers fundamental statistical analysis. Moreover, its outputs can be further analyzed with Ulamdyn.[7] Skitten was written for high performance, with minimum disk writing and OpenMP parallelization.

Skitten will be incorporated into and distributed with future versions of the Newton-X platform. For the moment, it can be obtained by directly contacting us.

## SM-4. Spin-Boson Hamiltonian Models

We tested SLED on two Spin-Boson Hamiltonian (SBH) models[8] with 5 (SBH-5) and 10 (SBH-10) dimensions. Their parametrization is described below. For MCTDH simulations, we worked in diabatic representation, while for SLED, Ehrenfest, FSSH, DC-FSSH, and AIMS, we employed their adiabatic version (ASBH).

The functional forms of diabatic energies and couplings, as well as of the adiabatic energies, energy gradients, and nonadiabatic coupling, are given in Table 1. The Wigner distribution used to sample initial conditions is also provided in the Table. All these quantities are explained in the Supporting Information of Refs. [6, 9]

Table 1. SBH models in diabatic and adiabatic representation with $N$ dimensions. For details, see the Supporting Information of Refs. [6, 9]

| Property | Equation |
|---|---|
| *Diabatic representation* | |
| Potential energy | $V_a = -\varepsilon_0 + \frac{1}{2}\sum_{j=1}^{N} M_j \omega_j^2 q_j^2 - \sum_{j=1}^{N} g_j q_j$ |
| | $V_b = \varepsilon_0 + \frac{1}{2}\sum_{j=1}^{N} M_j \omega_j^2 q_j^2 + \sum_{j=1}^{N} g_j q_j$ |
| Coupling | $V_{ab} = V_{ba} = -v_0$ |
| Lowest-state minimum | $q_k^{a,\min} = \dfrac{g_k}{M_k \omega_k^2} \quad (k = 1 \cdots N)$ |
| Wigner distribution around the lowest-state minimum | $P_W\left(q_k^a, p_k^a\right) = \dfrac{1}{\pi h}\exp\left(-\dfrac{M_k \omega_k}{h}\left(q_k^a - q_k^{a,\min}\right)^2\right)\exp\left(\dfrac{-1}{hM_k\omega_k}\left(p_k^a\right)^2\right)$ |
| *Adiabatic representation* | |
| Potential energy | $E_i = \frac{1}{2}\sum_{j=1}^{N} M_j \omega_j^2 q_j^2 + (-1)^i\left[\eta^2 + v_0^2\right]^{1/2} \quad (i = 1,2)$ |
| | $\eta = \left(\sum_{j=1}^{N} g_j q_j + \varepsilon_0\right)$ |
| Potential energy gradient | $\dfrac{\partial E_i}{\partial q_k} = M_k \omega_k^2 q_k + (-1)^i g_k\left[\dfrac{\eta}{\left[\eta^2 + v_0^2\right]^{1/2}}\right] \quad (i = 1,2; k = 1 \cdots N)$ |
| Nonadiabatic coupling | $F_{12}^k = -F_{21}^k = -\dfrac{1}{2}g_k\left[\dfrac{v_0}{\eta^2 + v_0^2}\right]$ |

## SBH-5

We designed a two-state Spin-Boson Hamiltonian (SBH) model[8] with five harmonic bath modes, each with a mass of $M_j = 1$ amu. In the isolated system, the two states have a half-energy separation of $\varepsilon_0 = 0.015$ a.u. with an inter-state diabatic coupling of $v_0 = 0.02$ a.u. The five harmonic baths are linearly coupled with the two-state system, where the coupling strengths are determined by the Debye spectral density with a highest cut-off frequency of 3000 cm⁻¹ and the reorganization energy of 0.075 a.u. Table 2 lists the harmonic frequencies and the coupling constants derived from these parameters. Figure 2 depicts the one-dimensional cuts of the potential energy surfaces along each mode, assuming the values of the other modes are zero.

Table 2: The discretized frequencies and coupling constants of the SBH-5 model.

| Bath Modes | $\omega_j$ (cm$^{-1}$) | $g_j$ (a.u.) |
|---|---|---|
| 1 | 255.138642 | 0.00383294 |
| 2 | 545.806988 | 0.00819964 |
| 3 | 930.527605 | 0.01397928 |
| 4 | 1554.795850 | 0.02335764 |
| 5 | 3000.00 | 0.04506889 |

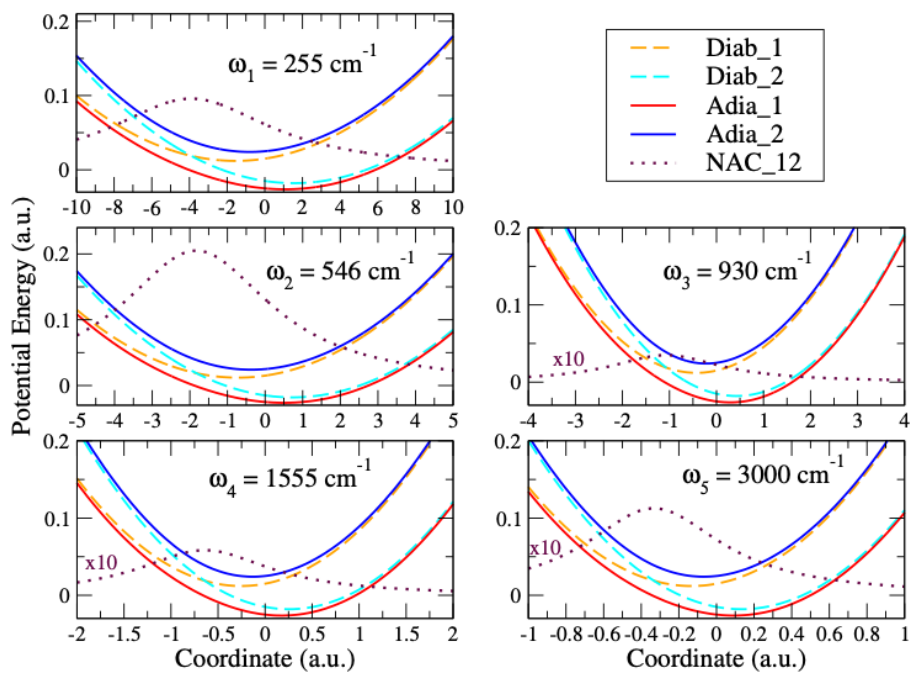

Figure 2: Potential energy curves of the SBH-5 model along each coordinate.

**SBH-10**

For convenience, we describe here the SBH model with ten harmonic bath modes, as introduced in Ref. [10] Each mode has a mass of $M_j = 1$ amu. In the isolated system, the two states have a half-energy separation of $\varepsilon_0 = 0.02$ a.u. with an inter-state diabatic coupling of $\nu_0 = 0.001$ a.u. The frequencies and coupling strengths listed in

Table 3 are determined by the Debye spectral density with a highest cut-off frequency of 3100 cm$^{-1}$ and the reorganization energy of 0.0125 a.u. Figure 3 depicts the one-dimensional cuts of the potential energy surfaces along each mode, assuming the values of the other modes are zero.

Table 3: The discretized frequencies and coupling constants of the SBH-10 model.

| Bath Modes | $\omega_j$ (cm⁻¹) | $g_j$ (a.u.) |
|---|---|---|
| 1 | 129.743484 | 0.000563 |
| 2 | 263.643268 | 0.001143 |
| 3 | 406.405874 | 0.001762 |
| 4 | 564.000564 | 0.002446 |
| 5 | 744.784527 | 0.003230 |
| 6 | 961.545250 | 0.004170 |
| 7 | 1235.657107 | 0.005358 |
| 8 | 1606.622430 | 0.006966 |
| 9 | 2157.558683 | 0.009356 |
| 10 | 3100.000000 | 0.013444 |

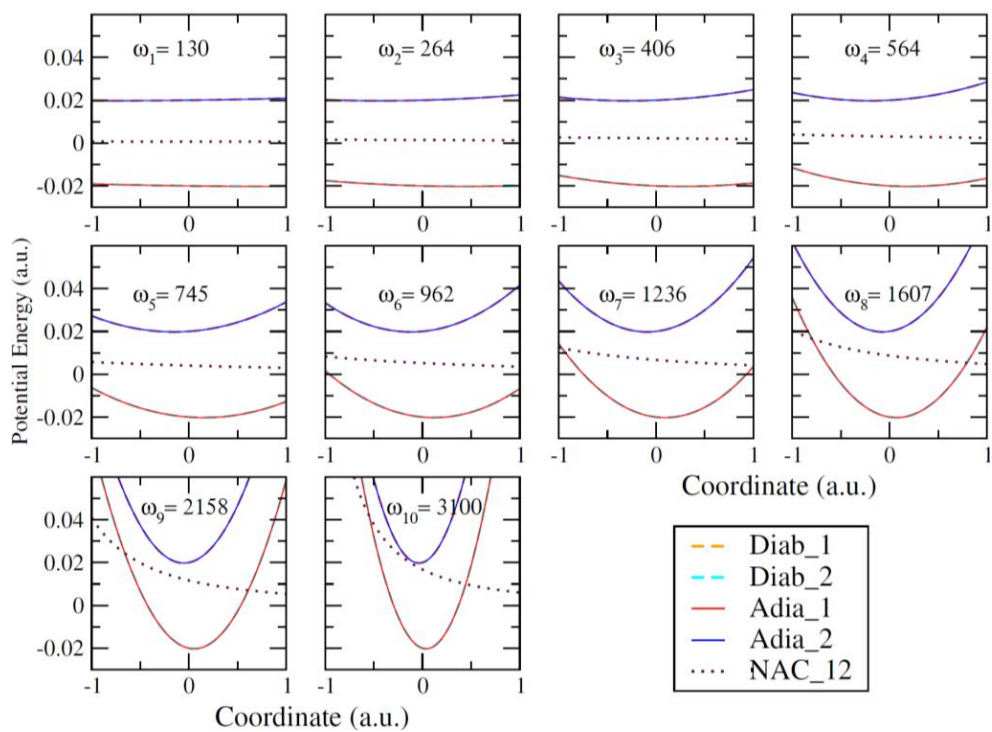

Figure 3: Potential energy curves of the SBH-10 model along each coordinate.

## SM-5. Quantum Dynamics of Tully Models

We performed quantum wave packet dynamics simulations on three one-dimensional Tully models,[4] simple avoided crossing (Tully 1), dual avoided crossing (Tully 2), and the extended coupling (Tully 3) model, proposed by Tully, using the Heidelberg MCTDH package[11] version 8.6.2 (http://mctdh.uni-hd.de). The wave function was propagated numerically exactly using the "exact" keyword as implemented in the MCTDH software. It expands the wavefunction and the Hamiltonian on the full product of primitive basis functions. We used the Fast Fourier Transformation (FFT) collocation basis for the primitive basis functions with varying grid sizes to attain convergence. The FFT grid spanned the range from -40 to 40 a.u. in one-dimensional coordinate space. The Short Iterative Lanczos (SIL) integrator was used with a maximal order of 20 and an error tolerance of $10^{-10}$.

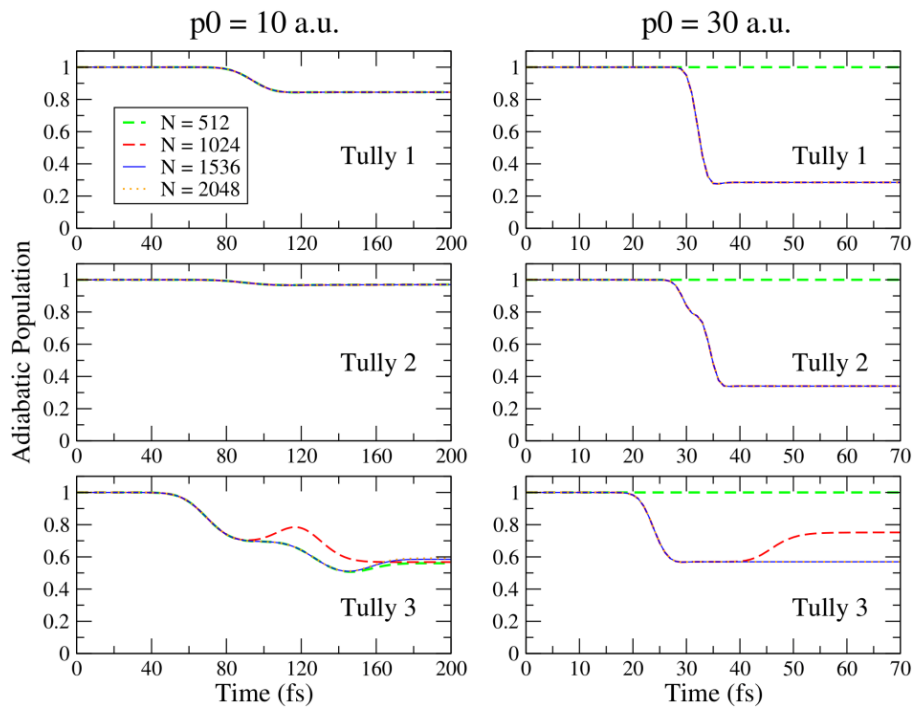

Figure 4: Convergence test of quantum dynamics simulation with respect to FFT grid size for the three one-dimensional Tully models.

In each quantum dynamics simulation, we placed a Gaussian wave packet at the lower diabatic state, centered at -20 a.u., with an initial momentum of either 10 a.u. or 30 a.u. to build the initial wave function. The width of the wave packet was set to $\Delta x = 20/p_0$, where $p_0$ denotes the initial momentum. The mass was set to 2000 a.u.

After simulating the wave packet dynamics in the diabatic representation, the adiabatic populations and the adiabatic coherence elements are calculated using the *adpop* analysis program inside the MCTDH package. We must acknowledge Prof. H. D. Meyer for providing a modified version of the *adpop* program to calculate the adiabatic coherence terms.

Figure 4 shows the convergence of the adiabatic population on the lower state with respect to the FFT grid size for two representative initial momenta (10 and 30 a.u.) for the three one-dimensional Tully models. Since the FFT grids with 1536 and 2048 points are the converged ones for all cases, we will only consider the quantum dynamics simulation for the one-dimensional models performed using the 1536-point FFT grid hereafter.

Thereafter, numerically exact quantum wavepacket dynamics simulations on the Tully 2 model were performed with a range of initial momenta, starting from 10 a.u. to 50 a.u., to monitor the lower adiabatic transmission probability as a function of initial momentum (see Figure 4 in the main text).

## SM-6. Benchmark dynamics of Tully models

Besides the quantum dynamics described in SM-5, we benchmarked the Tully models against several other established methods, including fewest-switches surface hopping (FSSH), decoherence-corrected fewest-switches surface hopping (DC-FSSH), Ehrenfest, and ab initio multiple spawning (AIMS).[12, 13]

For AIMS, we propagated the dynamics with a timestep of 0.05 fs. The overlap criterion was set low enough as not to affect spawning. The value for most systems was set to 0.7, with the only exception being Tully 2 with $p_0$ = 10 a.u., which was set to 0.2. Centroid energies and couplings were computed using the Saddle Point Approximation.[14] The coupling threshold of the NAC vector's norm was 0.08 a.u. for Tully 1, 0.01 a.u. for Tully 2, and 0.008 a.u. for Tully 3. Each dynamics used a total of 1000 initial conditions.

For FSSH and DC-FSSH, the same initial conditions were used, but each initial condition was repeated five times with different random seeds. The parameters are the same as in the main text, except for the timestep used in this case: 0.05 fs for the nuclear coordinates and 0.0025 fs for the electronic coefficients.

Lower-state population and coherence between lower and higher states for the three Tully models are shown in Figure 5 for high and low initial momentum, computed with FSSH, DC-FSSH, Ehrenfest, AIMS, and quantum dynamics.

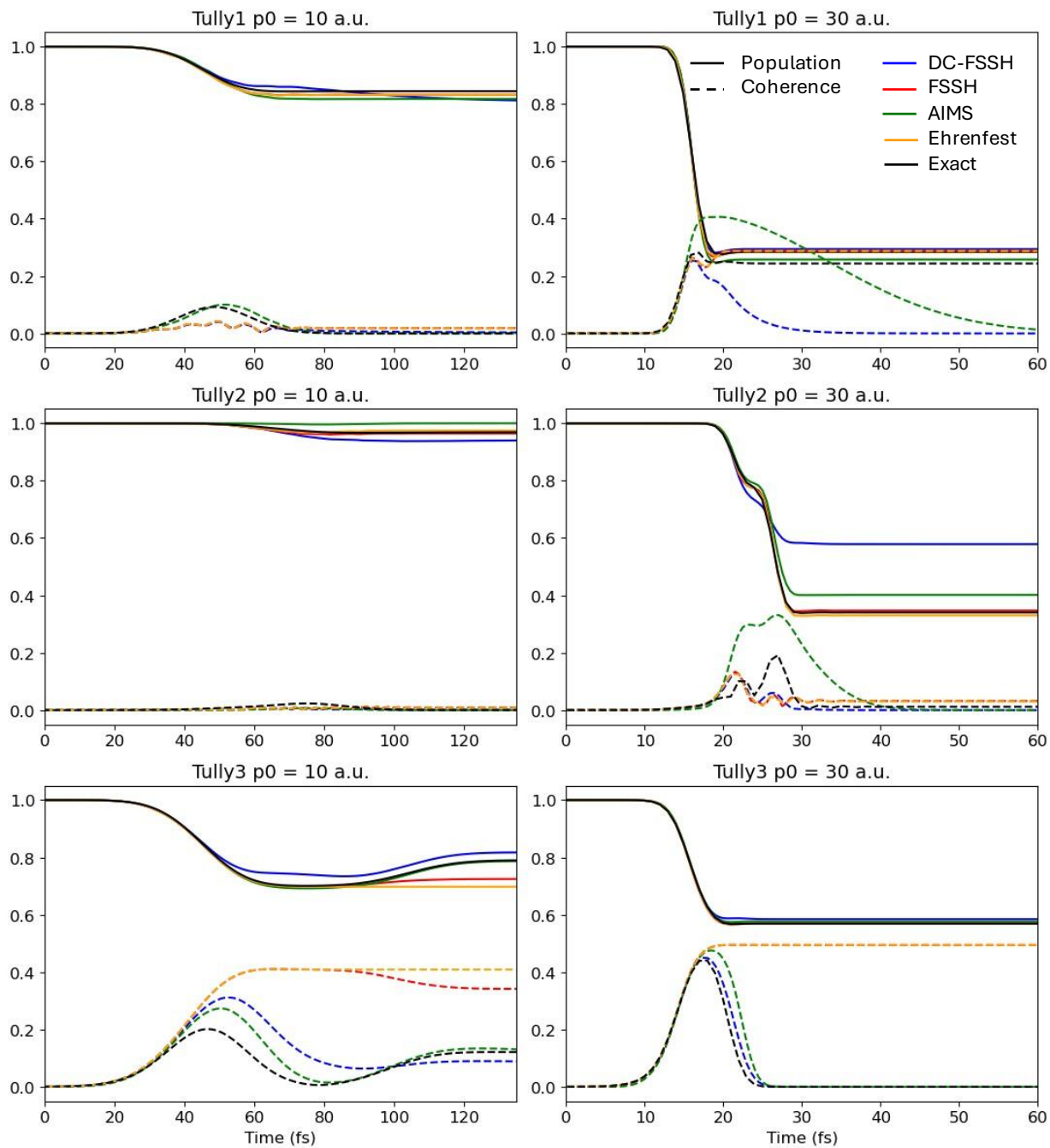

Figure 5. Population of the lower state (solid lines) and coherences between the lower and higher states (dashed lines) for Tully models 1 (top), 2 (middle), and 3 (bottom), computed with initial momentum 10 a.u. (left) and 30 a.u. (right), for several established methods.

## SM-7. Quantum Dynamics of the SBH-5 Model

We performed quantum wave packet dynamics on the SBH-5 model using the Heidelberg MCTDH package.[6] The multi-configurational time-dependent Hartree expansion is employed to represent the total wavefunction as a linear combination of time-dependent Hartree products, each constructed from single-particle functions that evolve in time. The Hermite DVR is used for the primitive basis, and the Harmonic Oscillator (HO) eigenfunctions are chosen as the initial wave function.

The initial wave packet is placed at the minima ($q_{k,min} = g_k/M_k\omega_k^2$) of the second diabatic state with an initial momentum of $p_k = \sqrt{\hbar M_k \omega_k}$, where $\omega_k$ and $g_k$, and $M_k$ are the frequency, coupling, and mass of the $k$-th mode, respectively (see SM-4). The wave packet is launched from the first diabatic state and propagated up to 1000 fs with an integration time step of 0.1 fs. The constant mean-field (CMF) integration scheme with an initial step size of 0.1 fs and an error tolerance of $10^{-8}$ is used for all MCTDH calculations. The MCTDH coefficients ($A$ vector) and SPFs are propagated by the short iterative Lanczos (SIL) algorithm (integration order 20 and an error tolerance of $10^{-10}$) and Bulirsch-Stoer (BS) (integration order 10 and error tolerance of $10^{-8}$) integrators, respectively. We have conducted systematic convergence tests by varying the number of single-particle functions (SPFs), $n$, and the grid size of the primitive basis functions, $N$.

Figure 6 shows the diabatic population decay curves as a function of time, varying the number of SPFs and the primitive basis grid size. Notably, the simulations achieved reasonable convergence by systematically increasing the basis size. However, to compare with other methods, we need the adiabatic populations from the diabatic wave packet propagation. This is not a trivial task, but rather a relatively complicated and costly one. Hence, we only calculated the adiabatic populations for the following simulations: ($n = 12, N = 12$), ($n = 12, N = 24$), and ($n = 16, N = 16$).

Figure 7 compares the adiabatic population of the second adiabat as a function of time obtained by DC-FSSH and quantum wave packet dynamics simulations. Despite being an approximate theory, the DC-FSSH result matches well with the MCTDH quantum dynamics simulation with a reasonable grid size ($n = 12, N = 24$). Hereafter, we only consider this one ($n = 12, N = 24$) for further comparison.

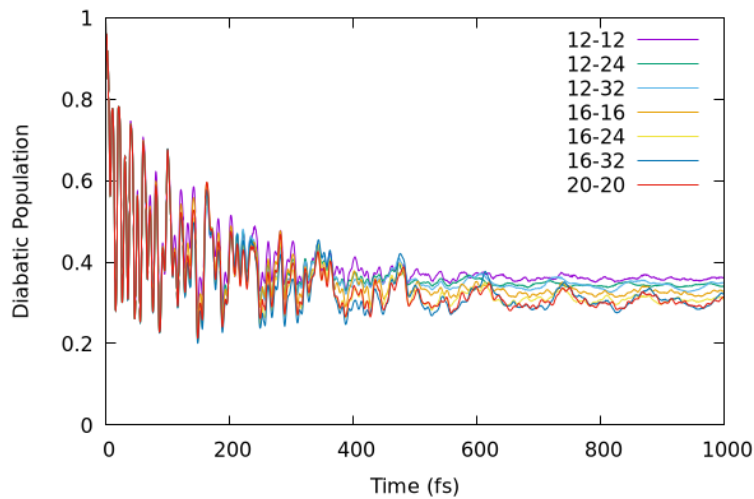

Figure 6: The diabatic population curves of the first diabatic state for different MCTDH simulations. Each curve is marked by two numbers: the first digit is the number of SPFs, and the second one denotes the number of primitive grid points.

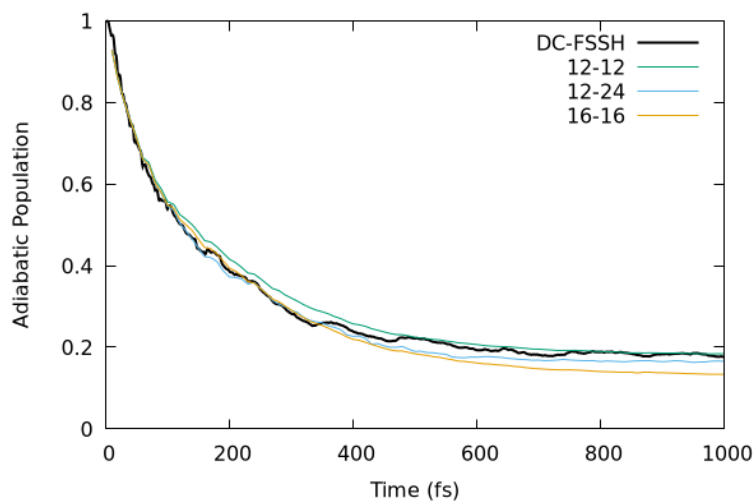

Figure 7: Comparison of the adiabatic population decay curves of the second adiabatic state between DC-FSSH and different MCTDH simulations. Each MCTDH simulation is marked by two numbers: the first digit is the number of SPFs, and the second one denotes the number of primitive basis functions.

## SM-8. Surface Hopping Dynamics of the SBH-5 Model

We have carried out the energy-based decohrence-corrected fewest-switch surface hopping (DC-FSSH) simulations (decoherence constant 0.1 a.u.) on the five-dimensional SBH model in the adiabatic representation (ASBH-5). 5000 independent trajectories are propagated up to 1000 fs with a timestep of 0.01 for the classical equation of motion. The coordinates and velocities for 5000 initial conditions are randomly sampled from uncorrelated (Wigner) normal distributions, as explained in Ref.[9] (see also SM-4). The energy distributions for the 5000 initial conditions are shown in Figure 8.

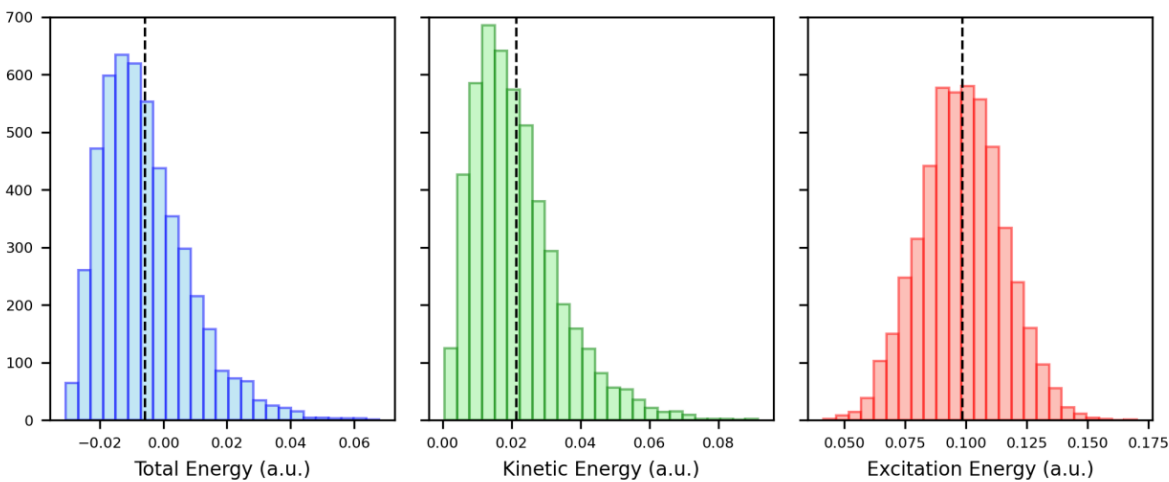

Figure 8: The distribution of total energy, kinetic energy, and the vertical excitation energy obtained from 5000 initial conditions used in the DC-FSSH simulation of the 5-dimensional SBH model. The dashed lines indicate the mean value of the distribution.

## Supplementary References